\documentclass[aps,notitlepage,superscriptaddress,prl]{revtex4-2}
\usepackage{ucs}
\usepackage[utf8x]{inputenc}
\usepackage{amsmath}
\usepackage{amsfonts}
\usepackage{amssymb}
\usepackage{amsthm}
\usepackage{fontenc}
\usepackage[dvips,pdftex]{graphicx}
\usepackage{setspace}
\usepackage{color}
\usepackage[dvipsnames]{xcolor}
\usepackage{graphicx}

\usepackage{multirow}
\usepackage{array}
\usepackage{subcaption}
\usepackage{tikz}
\usepackage{multimedia}
\usepackage{float}
\usepackage{rotating}
\usepackage{epstopdf}
\usepackage{mathtools}

\makeatletter
\@addtoreset{chapter}{part}
\makeatother

\newcommand{\blu}[1]{\textcolor{black}{#1}}
\newcommand{\dt}{\,\text{d}}

\newcommand{\pd}{\partial}
\newcommand{\Tr}{\text{Tr}}

\graphicspath{{C:/Users/Doron/Documents/PhD_work/Papers_And_Notes/Paper_Pics/Spherical_Ribbon/},{/home/doron/Documents/PhD_work/Papers_And_Notes/Paper_Pics/Spherical_Ribbon/}}

\begin{document}
	\title{Predicting the mechanical properties of spring networks}
	\date{\today}
	
	\author{Doron \surname{Grossman}}
	\email[]{doron.grossman@ladhyx.polytechnique.fr}
	\affiliation{LadHyX, CNRS, Ecole polytechnique, Institut Polytechnique de Paris, 91128 Palaiseau Cedex, France}
	\author{Arezki \surname{Boudaoud}}
		\affiliation{LadHyX, CNRS, Ecole polytechnique, Institut Polytechnique de Paris, 91128 Palaiseau Cedex, France}
	\date{\today}
		\begin{abstract}
		The elastic response of mechanical, chemical, and biological systems is often modeled using a discrete arrangement of Hookean springs, either representing finite material elements or even the molecular bonds of a system. However, to date, there is no direct derivation of the relation between a general discrete spring network\blu{, with arbitrary geometry,} and it's corresponding elastic continuum.  Furthermore, understanding the network's mechanical response requires simulations that may be expensive computationally.  Here we report a method to derive the exact elastic continuum model of any discrete network of springs, requiring network geometry and topology only. We identify and calculate the so-called "non-affine" displacements. Explicit comparison of our calculations to simulations of different crystalline and disordered configurations, shows we successfully capture the mechanics even of auxetic materials.  Our method is valid for residually stressed systems with non-trivial geometries, and is an essential step in generalizing active stresses on such discrete systems. It is easily generalizable to other discrete models, and opens the possibility of a rational design of elastic systems.
	\end{abstract}
	\maketitle

Since the 19th century \cite{germain1821recherches,cauchy1826exercises}, the theory of elasticity has \blu{mostly} been phenomenological. \blu{First principles derivations of elasticity - derivations of continuum elastic properties from known microscopic characteristics -  began to appear in the past 50 years.} \cite{Martin1972,Miserez2022,Lemaitre2006,Zaccone2023,Schlegel2016,Cui2019,Zaccone2011,Seung1988,lloyd2007identification,plummer2020buckling,sheinman2012nonlinear}. \blu{To date, past works have been focused on flat, compatible systems that do not exhibit residual stresses, and are difficult to generalize, even when rich non-linearities are introduced such as those stemming from complex topology\cite{sheinman2012nonlinear}, singular defects in crystalline order \cite{Seung1988} or active stresses\cite{scheibner2020odd,sheinman2012actively}}. \blu{Nevertheless}, it is widely accepted that in essence, elasticity may be described using spring-like interactions between constituents (e.g. first order approximation of intermolecular forces around equilibrium). In fact, elasticity is often described using a spring network either for computational or analytical aspects \cite{Gompper1996,Kot2014,gelder1998approximate} in a plethora of different systems and cases -  from modeling the shape of self assembled membranes \cite{Kantor1987,Seung1988,Bailey2021,Underhill2004,leembruggen2023computational,plummer2020buckling}, through biological systems \cite{Moshe2018,Chen2014}, to  modeling crack propagation \cite{Bolander1998} and various bio-inspired, and meta materials \cite{Rocks2017,Liu2019,Luo2023,zhang2024stretchable}. Typically, calculation of the network's elastic response can only be done via direct simulation of a loading scheme (i.e. simulating a \blu{specific} mechanical load and the response to it).

In this paper we directly derive a generalized elastic continuum limit of any triangulated spring network, with arbitrary reference lengths and spring constants, in two and three dimensions\blu{, within arbitrary geometry, including residually stressed systems}. The resulting continuum limit depends solely on the network geometry and topology, as expressed by reference lengths, spring constants, and bonds. From this description, any macroscopic elastic quantity can be extracted, such as Poisson's ratio. We demonstrate the strength of this approach, by calculating Poisson's ratios for different test cases, both ordered and disordered, recovering even auxetic behavior. We identify and directly calculate the so-called "non-affine" displacement \blu{of every element}. These are local deformations deviating from the local average deformation, and are responsible of the wide range of responses seen in disordered elastic media. The results are valid for residually stressed elastic systems {by virtue of the mathematical formulation setting of the problem}. 

The continuum limit we derive is formulated within the theory of incompatible elasticity,\cite{Efrati2009} which is a modern formulation of elasticity, that successfully describes residually stressed elastic systems \cite{Zhang2019,Levin2021,Efrati2013}. In this formulation, an elastic material is described by a metric $\mathbf{g}$ with elements $g_{\mu\nu}$, which describes actual distances between material elements, and a reference metric $\mathbf{\bar{g}}$ with elements $\bar{g}_{\mu\nu}$ describing ideal distances. The elastic energy then depends on the squared difference of $\mathbf{g}- \mathbf{\bar{g}}$, $E_{el} \propto \left\|\bf{g -\bar{g}}\right\|^2$, for some proper choice (yet to be defined) of the norm $\| \cdot \|^2$, through the elastic (four-indexed) tensor  $\bar{\bar{A}}$ (with elements $A^{\mu\nu\alpha\beta}$).

This description, via use of metrics, is very similar in essence and form to the classical description of Hookean springs. It is independent from assumptions about the existence of a rest configuration, which enables the treatment of residual stresses, \blu{ and distinguishes this framework from other approaches to elasticity} .  
In the following, we will consider a discrete network of springs and show how such formulation naturally arises. We will then coarse grain the network and will identify the non-affine quantities, show how they contribute to the elastic continuum energy\blu{, and shortly discuss how this derivation relates to other methods found in the literature}.

\section{Results}
\subsection{Framework}
The theory of incompatible elasticity \cite{Efrati2009} is the framework to which the results of this paper are anchored to. Within it, the elastic energy is given by:

\begin{align}\label{eq: incompatible energy}
	E_{el} = \int A^{\mu\nu\alpha\beta}\left(g_{\mu\nu}- \bar{g}_{\mu\nu}\right)\left(g_{\alpha\beta}- \bar{g}_{\alpha\beta}\right) \dt V_{\bar{g}}
\end{align}

Where $g_{\mu\nu}$ is the actual metric, describing distances between neighboring material points, $\bar{g}_{\mu\nu}$ is the reference metric, describing ideal distances. $\dt  V_{\bar{g}} = \sqrt{\bar{g}} d^Mx$ is the volume element in $M$ dimensions, $\bar{g}= \det{\bf \bar{g}}$. $A^{\mu\nu\alpha\beta}$ are the elements of the elastic tensor. Einstein' summation is assumed for repeating upper and lower Greek indices.  Greek indices refer to coordinates within the volume of the $n$ dimensional manifold.

In an isotropic material, $A^{\mu\nu\alpha\beta} = \frac{Y}{16(1-\nu^2)} \left[\frac{1}{2}(1-\nu)\left(\bar{g}^{\mu\alpha}\bar{g}^{\nu\beta}+\bar{g}^{\nu\alpha}\bar{g}^{\mu\beta}\right)+ \nu \bar{g}^{\mu\nu}\bar{g}^{\alpha\beta}\right]$, where $\bar{g}^{\mu\nu}$ is the inverse reference metric, $Y$ is Young's modulus, setting the rigidity scale of the system, and $\nu$ is Poisson's ratio, describing the amount a material contracts in one axis, when the other is stretched (negative values indicate expansion). In non isotropic materials Poisson's ratio is orientation dependent, and the expression of $A^{\mu\nu\alpha\beta}$ will typically depend on additional terms.

The elastic stress is given by the variation $\sigma^{\mu\nu}= \frac{\delta E_{el}}{\delta g_{\mu\nu}}$, and the material satisfies the usual force balance equation:\begin{align} \label{eq:elastic equation}
\bar{\nabla}_\mu \sigma^{\mu\nu} +\sigma^{\mu\alpha}\left(\Gamma_{\mu\alpha}^\nu-\bar{\Gamma}_{\mu\alpha}^\nu\right) = f_{ext}^\nu
\end{align}  where $f_{ext}^\nu$ are the external forces acting on the systems, and $\sigma^{\mu\nu} = \frac{\delta E_{el}}{\delta g_{\mu\nu}}$ is the elastic stress, $\bar{\nabla}_{\mu}$ is the covariant derivative with respect to $\bar{g}_{\mu\nu}$ and $\Gamma^\alpha_{\beta\gamma},\bar{\Gamma}^\alpha_{\beta\gamma}$ are the christoffel symbols associated with the metrics $g_{\mu\nu}$ and $\bar{g}_{\mu\nu}$ respectively. \blu{The second term on the left-hand-side of the equation corresponds to the contribution of residual stresses to the total stresses in the system.}

\subsection{Analytical Derivation}
\begin{figure}[!h]
		\centering
		\includegraphics*[width= .4\textwidth]{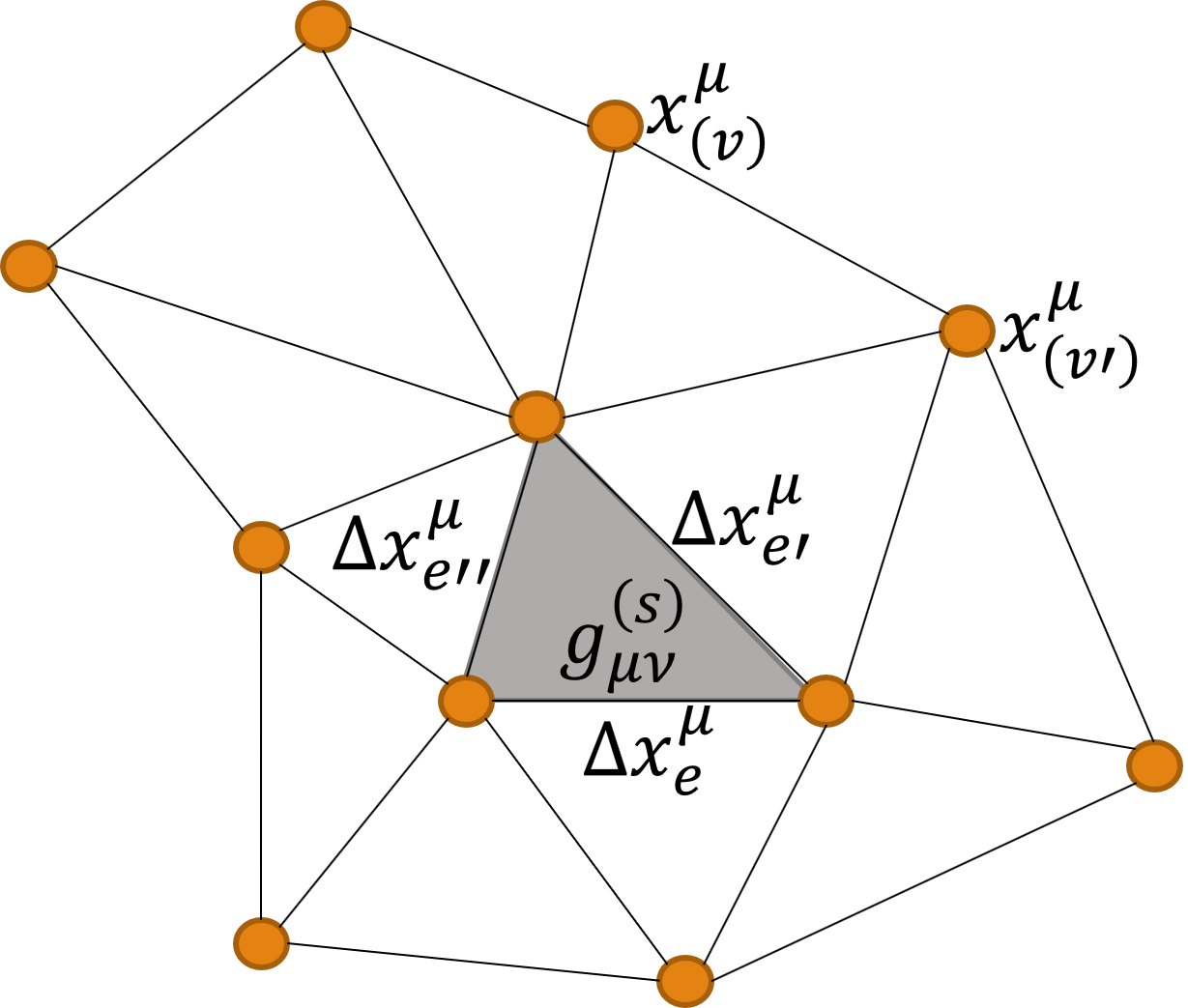}
		\caption{Depiction of a small portion of spring network. Showing the coordinates of different vertexes ($x^
			\mu$), the edge vectors $\Delta  x^\mu$ of the edges $\{e,e',e''\}$ associated with triangle ($s$), and the local metric  $g^{(s)}_{\mu\nu}$ (shaded triangle).  } \label{fig: definitions}
	\end{figure}

We begin by considering  a triangulated mesh of springs, each with reference length $\bar\ell_e$, spring constant $k_e$ and an actual length $l_e$, where the index $e$ enumerates the springs. The elastic energy of the systems is exactly given by 	
	\begin{align}
		E_{el} = \sum_e \frac{1}{2}k_e \left(l_e - \bar\ell_e\right)^2.
 		\end{align}
 	A triangulated network is easily divided into sum of specific simplexes (cells). In three dimensions these are tetrahedrons, and in two dimensions these are simple triangles. {The energy can then be re-written as a sum over simplexes, and taking care not to count the same edge twice}	
	\begin{align}
		E_{el} = \frac{1}{2}\sum_s  \sum_{e \in s} \frac{1}{2}k_e \left(l_e - \bar\ell_e\right)^2.
	\end{align}
	Here, the index $s$ enumerates simplexes, $e \in s$ means summation over all the springs associated with the simplex $s$.
	When left to relax, the network assumes some configuration (not necessarily unique) $\{\vec{f}_v\}$ in $\mathbb{R}^n$ for every vertex $v$.
	By Setting coordinates $x^\mu_v$ to each vertex $v$ ($\mu$ is the coordinate component), given actual lengths $\{l_e\}$,  we may uniquely define a "local metric" 	$g^{(s)}_{\mu\nu}$ associated with a cell $s$, so that 
	\begin{align}\label{eq: local_metric}
	l_e^2 = g^{(s)}_{\mu\nu} \Delta x^\mu_e \Delta x^\nu_e ~~~ \forall 	e \in s.
	\end{align} Where $\Delta x^\mu_{(1,2)} = x_2^\mu - x_1^\mu$ is the "coordinate difference" of the edge $e=(1,2)$ connecting vertexes $1$ and $2$.  \eqref{eq: local_metric} is not an approximation, rather it is an exact definition of the local quantity $g^{(s)}_{\mu\nu}$, over the whole simplex (see Fig. \ref{fig: definitions}). 
	
	 Hence, a given simplex uniquely defines a local metric, $g^{(s)}_{\mu\nu}$, associated to it. A physical systems is constrained such that any two local metrics $\mathbf{g^{(i)}}$ and $\mathbf{g^{(j)}}$ with a shared edge $\Delta x_e$, agree on its length- $l_e[\mathbf{g}^{(i)}] = l_e[\mathbf{g}^{(j)}]$, where $l_e[\mathbf{g}]$ is the edge's length, as measured using the metric $\mathbf{g^{}}$. 
	We can now rewrite the energy -
	\begin{align}\label{eq: no ref metric}
		E_{el} = \frac{1}{4} \sum_s\sum_{e \in s} k_e \left(\sqrt{{g^{(s)}}_{\mu\nu} \Delta x_e^\mu \Delta x_e^\nu} - \bar\ell_e\right)^2.
		\end{align}

In order to advance, we introduce three assumption. First we assume that the reference lengths $\{\bar\ell_e\}$ are compatible, so that a single simplex can assume the shape described by the lengths $\bar\ell_e$. {This assumption definitely holds in any non-residually stressed systems, curved or not, but in fact also most residually stressed systems. It might fail in certain glasses (see discussion for more) }.  This assumption means that the reference lengths locally define a reference metric ${\bar{g}^{(s)}}_{\mu\nu}$. Under this assumption we can write $\bar\ell_e = \sqrt{\bar{g}^{(s)}_{\mu\nu}\Delta x_e^\mu \Delta x_e^\nu}$.

Second, we assume that in an equilibrium configuration (again, not necessarily unique), deviations of actual lengths from the reference lengths are small. {This is true even for most residually - stressed systems. As the curvatures are typically on a much larger scale  than the element scale (i.e atomic/molecular or even mesoscopic).}
\begin{align}
	l_e - \bar\ell_e = \frac{l_e^2 - \bar\ell_e^2}{l_e +\bar\ell_e} \simeq \frac{1}{2 \bar\ell_e} \left(l_e^2 - \bar\ell_e^2\right) + \cdots
\end{align}
$\cdots$ marks higher order terms of $l_e^2 -\bar\ell_e^2$. 
Thus {exchanging the expressions for $l_e$ and $\bar\ell_e$ with their $g$ and $\bar{g}$ expressions we get}
\begin{align}
	E_{el} =  \sum_s\sum_{e \in s} \frac{k_e}{16 \bar\ell_e^2} \left(g^{(s)}_{\mu\nu} \Delta x_e^\mu \Delta x_e^\nu - \bar{g}^{(s)}_{\mu\nu} \Delta x_e^\mu \Delta x_e^\nu\right)^2 + \cdots
\end{align}
We expand the energy:
\begin{align}
	E_{el} = &  \sum_s \left(g^{(s)}_{\mu\nu}-\bar{g}^{(s)}_{\mu\nu}\right) \left(g^{(s)}_{\alpha\beta}-\bar{g}^{(s)}_{\alpha\beta}\right) \sum_{e \in s} \frac{k_e \Delta x_e^\mu \Delta x_e^\nu \Delta x_e^\alpha \Delta x_e^\beta}{16 \bar\ell_e^2} .
\end{align}
Marking the local elastic tensor \begin{equation} \label{eq: A simp}
	A_{(s)}^{\mu \nu \alpha \beta} = \sum_{e \in s} \frac{k_e \Delta x_e^\mu \Delta x_e^\nu \Delta x_e^\alpha \Delta x_e^\beta}{16 \bar\ell_e^2},
\end{equation} we write -
\begin{align}\label{eq: discrete energy}
	E_{el} = &  \sum_s A_{(s)}^{\mu\nu\alpha\beta} \left(g^{(s)}_{\mu\nu}-\bar{g}^{(s)}_{\mu\nu}\right) \left(g^{(s)}_{\alpha\beta}-\bar{g}^{(s)}_{\alpha\beta}\right) 
\end{align}
This equation is  very similar to \eqref{eq: incompatible energy}, and may be considered as a discrete version of that equation. 

The {third and} last assumption introduced, is that $\bar{g}_s$ varies slowly on some large enough region. Without this assumption the continuum limit cannot hold (though an effective continuum may be derived, in principle). {As before, this is very common in continuous media, where the actual elements are very small relative to almost any other quantity (see discussion for more)}.

\begin{figure}
	\centering
	\includegraphics[width=0.5 \textwidth]{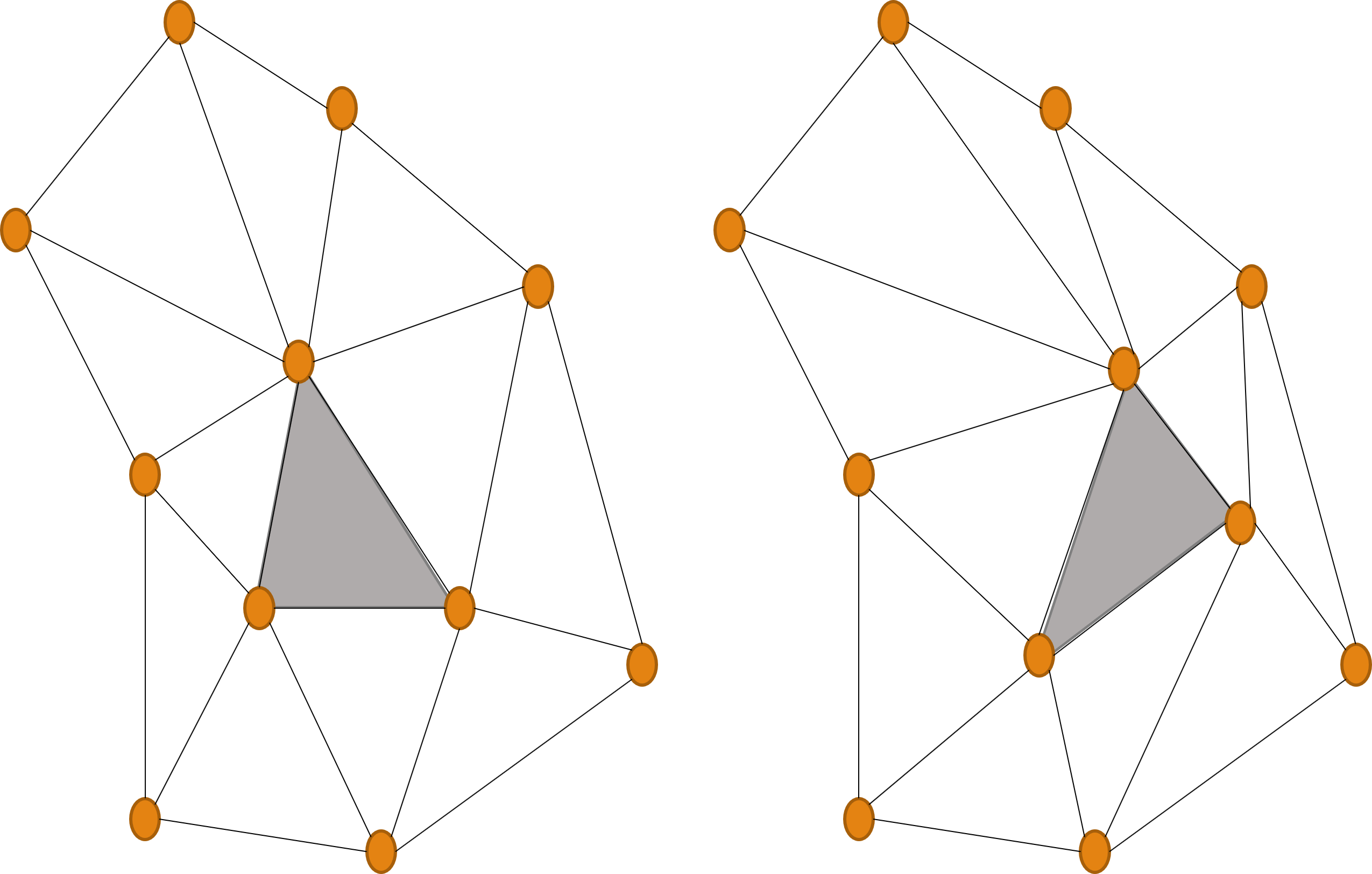}
	\caption{Schematic (exaggerated) of non-affine deformations. The same network from Fig. \ref{fig: definitions} stretched along the $y$ axis. On the left - the network when only affine deformation are allowed, every element is elongated by the same ratio. On the right - a response with non affine deformations. Different element may respond differently to the stretch. Shaded triangle is for visualization only.\label{fig: non affine vs affine }}
\end{figure}

Defining an average metric, ${g}_{\mu\nu}$, on some neighborhood {$\Omega$}, we may expand $g^{(s)}_{\mu\nu}= {g}_{\mu\nu} + \delta g^{(s)}_{\mu\nu}$. Formally, $\delta g^{(s)}_{\mu\nu}$ describe the "non-affine" deformations (see Fig. \ref{fig: non affine vs affine }). The energy then reads -
\begin{align}
	E_{el}= \sum_\Omega \left( \sum_{s\in \Omega} A_{(s)}^{\mu\nu\alpha\beta} \Delta g_{\mu\nu}\Delta g_{\alpha\beta} + \sum_{s\in \Omega} A_{(s)}^{\mu\nu\alpha\beta}\Delta g_{\mu\nu}\delta g^{(s)}_{\alpha\beta} +\sum_{s\in \Omega} A_{(s)}^{\mu\nu\alpha\beta}\delta g_{\mu\nu}\Delta g^{(s)}_{\alpha\beta} + \sum_{s\in \Omega} A_{(s)}^{\mu\nu\alpha\beta} \delta g^{(s)}_{\mu\nu} \delta g^{(s)}_{\alpha\beta}  \right)
,\end{align}
where $\Delta g_{\mu\nu} = {g}_{\mu\nu} - \bar{{g}}_{\mu\nu}$, and $\sum_\Omega$ is the sum over all the neighborhoods in which $\mathbf{g}$ and $\mathbf{\bar{g}}$ may be regarded as constant. 
Since, under our assumptions, if ${g}_{\mu\nu} =  {\bar{g}}_{\mu\nu}$,  $\delta g^{(s)}_{\mu\nu} =0 \, \forall s$, then for small deviations from $\bar{{g}}_{\mu\nu}$, $\delta g^{(s)}_{\mu\nu} =  W_{(s)\,\mu\nu}^{\alpha\beta}\Delta g_{\alpha\beta}$. {The proportion tensors $ W_{(s)\,\mu\nu}^{\alpha\beta}$ describe how the local metric (basically the shape) of a triangle changes under a general deformation (described by the metric difference $\Delta g_{\mu\nu}$), with respect to the average  metric $g_{\mu\nu}$}. While mathematically different, we identify the proportion tensors $W_{(s)\,\mu\nu}^{\alpha\beta}$, with the "non affine" deformations of each simplex, which are yet unknown. {Note, that this linear expression does not  assume  that a situation where ${g}_{\mu\nu} =  {\bar{g}}_{\mu\nu}$ is indeed achievable (i.e - the system may be residually stressed \cite{Efrati2009}) and is only dependent on the assumption of small deformation.}

The elastic energy $E_{el}^\Omega$ within a single neighborhood ($E_{el}=\sum_\Omega E_{el}^\Omega$)   then reads
\begin{align}\label{eq: enerrgy_lagrange}
	E_{el}^\Omega = & \sum_{s} \left(A_{(s)}^{\mu\nu\alpha\beta} +  A_{(s)}^{\mu\nu\rho\sigma}W_{(s)\,\rho \sigma}^{\alpha \beta} +A_{(s)}^{\alpha\beta\rho\sigma}W_{(s)\,\rho \sigma}^{\mu \nu}+ A_{(s)}^{\tau\lambda\rho\sigma}W_{(s)\,\rho \sigma}^{\alpha \beta}W_{(s)\,\tau \lambda}^{\mu\nu}\right)\Delta g_{\mu\nu}\Delta g_{\alpha\beta} \\ \nonumber
	& + \chi^{\alpha\beta} \sum_{s}  W_{(s) \alpha\beta}^{\mu\nu}\Delta g_{\mu\nu}
\end{align}
The second line is a Lagrange term forcing the requirement that $\sum_{s}\delta g_{(s) \mu\nu} =0$.  
We note, that as $g_{\mu\nu}$ is an average metric, $\sum_s \delta g^{(s)}_{\mu\nu} = 0$. This translates to $\sum_{s} W_{(s)\, \mu\nu}^{\alpha\beta} =0$. 
\blu{Note,}{this is a simplified requirement, enough to \blu{calculate the} geometrical response of the system as is used throughout this paper. However, \blu{when calculating the energetic} (rigidity) response of the system, a different approximation is required \blu{(or else the exact constraints)}, see discussion in appendix E. \blu{Given that} in many applications \blu{we are interested in the geometric response}, we continue with this approximation. For exactness, the complete constraints \blu{ include the $\chi$ term above and} the requirement that neighboring "local metrics" agree on the same edge length.} 

{\blu{ We now seek} to find the elastic response of  \blu{the} system, that is  to find both $g_{\mu\nu}$ and all $ W_{(s)\,\mu\nu}^{\alpha\beta}$. To this end we take the functional derivative of the energy with respect to \blu{ $g_{\mu\nu}, W^{\alpha\beta}_{(s) \mu\nu}$, and $\chi^{\alpha\beta}$}  and find: } 
\begin{align}\label{eq: Field equations}
\bar{\nabla}_\mu \sigma^{\mu\nu} +\sigma^{\mu\alpha}\left(\Gamma_{\mu\alpha}^\nu-\bar{\Gamma}_{\mu\alpha}^\nu\right) = f_{ext}^\nu \\ \nonumber
\left(A_{(s)}^{\mu\nu\alpha\beta}+A_{(s)}^{\lambda\tau\alpha\beta}W_{(s)\,\lambda \tau}^{\mu\nu}\right) \Delta g_{\mu\nu} + \frac{1}{2} \chi^{\alpha\beta} =0
\end{align}
The first line is actually the elastic equation \eqref{eq:elastic equation}, {derived from the variation of the energy (Eq.\eqref{eq: enerrgy_lagrange}) with respect to the metric \cite{Efrati2009}, and the second line was obtained through variation of $\{W^{\mu\nu}_{(s) \alpha \beta}\}$}. At this point it is enough to note that $\sigma^{\mu\nu}= \frac{\delta E_{el}}{\delta g_{\mu\nu}}$, where $E_{el}$ is given by Eq. \eqref{eq: enerrgy_lagrange},  that $\Gamma^\alpha_{\beta\gamma}$ and $\bar\Gamma^\alpha_{\beta\gamma}$ are the Christoffel symbols associates with the metrics $\bf{g}$ and $\bf{\bar{g}}$, and $\bar{\nabla}_\mu$ is the covariant derivative associated with the metric $\bf{\bar{g}}$.

{Simplifying further, by summing the second line of Eq. \eqref{eq: Field equations} over all triangles, we find that $$\frac{1}{2} \chi^{\alpha\beta} = - \frac{1}{n} \sum_s \left(A_{(s)}^{\mu\nu\alpha\beta}+A_{(s)}^{\lambda\tau\alpha\beta}W_{(s)\,\lambda \tau}^{\mu\nu}\right) \Delta g_{\mu\nu} $$ where $n$ is the number of simplexes in the neighborhood $\Omega$. We can now rewrite the second equation in \eqref{eq: Field equations}, by defining {the naive elastic tensor $A^{\mu\nu\alpha\beta}$ (describing the "bare" elastic response without the non-affine displacements), the true elastic tensor $\tilde{A}^{\mu\nu\alpha\beta}$  (describing the actual response when non-affine deformations are present), and the local deviation from the naive elastic tensor $\delta A_{(s)}^{\mu\nu}$}:
	\begin{align}
			A^{\mu\nu\alpha\beta} &= \frac{1}{n} \sum_{s} A^{\mu\nu\alpha\beta}_{(s)} \\  
			\tilde{A}^{\mu\nu\alpha\beta} &=  \frac{1}{n} \sum_s \left(A_{(s)}^{\mu\nu\alpha\beta} +  A_{(s)}^{\mu\nu\rho\sigma}W_{(s)\,\rho \sigma}^{\alpha \beta} +A_{(s)}^{\alpha\beta\rho\sigma}W_{(s)\,\rho \sigma}^{\mu \nu}+ A_{(s)}^{\tau\lambda\rho\sigma}W_{(s)\,\rho \sigma}^{\alpha \beta}W_{(s)\,\tau \lambda}^{\mu\nu}\right)\\
			\delta A^{\mu\nu\alpha\beta}_{(s)} &= {A}^{\mu\nu\alpha\beta} _{(s)}- A^{\mu\nu\alpha\beta},
	\end{align} 
 marking $\langle A_{(s)} W_{(s)}\rangle^{\mu\nu\alpha\beta} = \frac{1}{n}\sum_{s} A_{(s)}^{\lambda\tau \alpha \beta} W_{(s)\,\lambda\tau}^{\mu\nu}$, and using  $\langle A_{(s)} W_{(s)}\rangle^{\mu\nu\alpha\beta} = \langle \delta A_{(s)} W_{(s)}\rangle^{\mu\nu\alpha\beta}$, we may rewrite the second line of \eqref{eq: Field equations}, after a little algebra
\begin{align}\label{eq: non affine}
\delta A_{(s)}^{\mu\nu\alpha\beta}	+ A_{(s)}^{\lambda\tau \alpha \beta} W_{(s)\,\lambda\tau}^{\mu\nu}-\langle \delta A_{(s)} W_{(s)}\rangle^{\mu\nu\alpha\beta}=0.
\end{align}}
This is a linear equation for the non-affine deformation terms, $W$. It is solved by mapping the tensor components and indexes unto a multi - index notation and using the symmetries of the tensors (in two dimensions $W_{(s)\, \mu \nu}^{\alpha\beta}$ has only 9 independent entries, while in three dimension it has 36)
\begin{align}
 \delta A_{S} + \sum_{S'} \left( A_{S S'} - B_{S S'}\right) W_{S'}=0\end{align}
Where $\delta A_S$, $A_{SS'}$ and $B_{SS'}$ are reorganizations of the elements of $\{A_s^{\mu\nu\alpha\beta}\}$ and $\{\delta A_s^{\mu\nu\alpha\beta}\}$ into matrices compatible with the new multi index (see appendix A in \cite{supp}). The solution -
\begin{align} \label{eq: non_affine_cental}
	 W_S = -\sum_{S'}{\left[A-B\right]^{-1}}_{SS'} \delta A_{S'}.
\end{align}
We may now write the elastic energy - 
\begin{align}\label{eq: energy_full_disc}
E_{el} = &  \sum_\Omega n \tilde{A}^{\mu\nu\alpha\beta} \left(g_{\mu\nu}-\bar{g}_{\mu\nu}\right) \left(g_{\alpha\beta}-\bar{g}_{\alpha\beta}\right). 
\end{align}
This equation is essentially already coarse grained - \begin{align}\label{eq: continuum full}
E_{el} = &  \sum_\Omega n \tilde{A}^{\mu\nu\alpha\beta} \left(g_{\mu\nu}-\bar{g}_{\mu\nu}\right) \left(g_{\alpha\beta}-\bar{g}_{\alpha\beta}\right) \frac{V^\Omega_\mathbf{\bar{g}}}{V^\Omega_\mathbf{\bar{g}}} = \sum_\Omega  \rho_\Omega\tilde{A}^{\mu\nu\alpha\beta} \left(g_{\mu\nu}-\bar{g}_{\mu\nu}\right) \left(g_{\alpha\beta}-\bar{g}_{\alpha\beta}\right) \int_\Omega \dt V_{\mathbf{\bar{g}}}    \\ \nonumber  =& \int \tilde{A}^{\mu\nu\alpha\beta} \left(g_{\mu\nu}- \bar{g}_{\mu\nu}\right)	\left(g_{\alpha\beta}- \bar{g}_{\alpha\beta}\right) \dt V_{\bar{g}}	
\end{align}
where $V^\Omega_\mathbf{\bar{g}}$ is the volume of the neighborhood $\Omega$, $\rho_\Omega=n/ V^\Omega_\mathbf{\bar{g}}$ is the local density (which we absorb into the definition of $\tilde{A}^{\mu\nu\alpha\beta}$),$\int_\Omega \dt V_\mathbf{\bar{g}}$ is an integral over the region $\Omega$ and we use the fact the $\sum_\Omega \int_\Omega \dt V_\mathbf{\bar{g}} = \int \dt V_\mathbf{\bar{g}}$ over all the network.

Eqs. \eqref{eq: continuum full}, and \eqref{eq: non_affine_cental} form the central result of this work. Together with the definitions of $A^{\mu\nu\alpha\beta},\tilde{A}_s^{\mu\nu\alpha\beta},W_{(s)\,\lambda\tau}^{\mu\nu}$, they fully describe the response of the network and offer a novel way of computing it directly from the network geometry, without the need to consider any specific load. Under this view  the metric $\mathbf{g}$  is the actual, coarse grained, metric of the system,  $\mathbf{\bar{g}}$ describes the reference geometry of the system, and $\tilde{A}^{\mu\nu\alpha\beta}$ is the coarse grained elastic tensor, governing the mechanical response (as opposed to the local or "bare" term $A^{\mu\nu\alpha\beta}_{(s)}$).  $W^{\mu\nu}_{(s) \, \alpha \beta}$ describe the non-affine displacements. 

\subsection{Comparison to simulation}
 Results were tested numerically by comparing the expected Poisson's ratio and Young's modulus using the above scheme, to that of simulated two dimensional triangulated spring networks. {The (possible orientation dependent ) Poisson ratio was chosen as a measure for comparison as this is the quantity describing the complex response of the system. In isotropic materials, there is only one other independent quantity- Young's modulus}. In general, we find a very good agreement between theory and simulation, the details of which are described in the methods section. We considered three cases - ordered, foam-like, and honeycomb networks.
 
 {We simulated a 4:1 aspect ratio ribbons with a given element geometry and minimized their elastic energy after clamping their far ends and stretching them slightly (see methods for exact details). The 4:1 ratio was chosen to eliminate boundary effects that result from the clamping. Energy was minimized using a simple gradient descent algorithm (from the SciPy Python package) over the position of each vertex. Width measurement was done at the middle of the ribbon as far from both clamped edges as possible.}

\subsubsection{Ordered netwroks}
In the ordered case we simulated a triangular lattice of varying unit cells' shapes, and computed the angle dependence of Poisson's ratio. A unit cell's shape was quantified using two parameters - the shear factor $\phi$ and elongation factor $\psi$ to quantify the deviation from an equilateral triangle (when $\phi=\psi=1$). In this case all of the non-affine tensors $W_{(s)\, \mu\nu}^{\alpha\beta} =0$ identically vanish, leading to a simple calculation using Eq.\eqref{eq: A simp} (detailed analysis in appendix C  \cite{supp}). In Figure \ref{fig: numerics_vs_analytics}, we see a comparison between the analytical solution, and the numerical estimation. Insets show the lattice structure, at  an arbitrary orientation chosen as $\theta =0$. {In general, good agreement is seen between theory and simulation. Some systematic errors might be seen near graph minima and maxima. This is especially true for subfigure b). These systematic improve on higher resolution and are related to finite size effects.}

\begin{figure}[h]
	\centering
	\includegraphics*[width=0.9\textwidth]{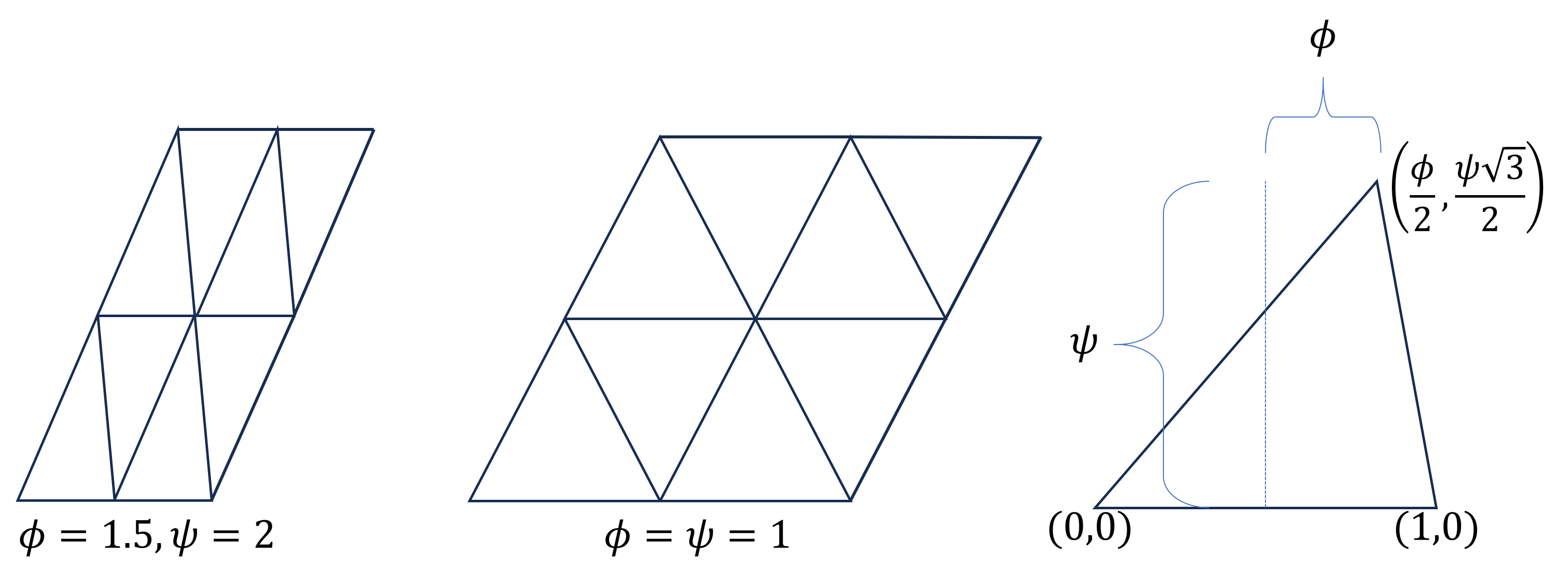}
	\caption{Quantifying the shape of unite cell, using the shear ($\phi$) and elongation ($\psi$) factors, as described in the right pictures. Coordinates are used to demonstrate how to create the triangle, up to scaling. On the left - demonstration of two possible lattices as indicated. \label{fig: triangles}}
\end{figure}

\begin{figure}[h]
	\centering
	\includegraphics*[width= .9\textwidth]{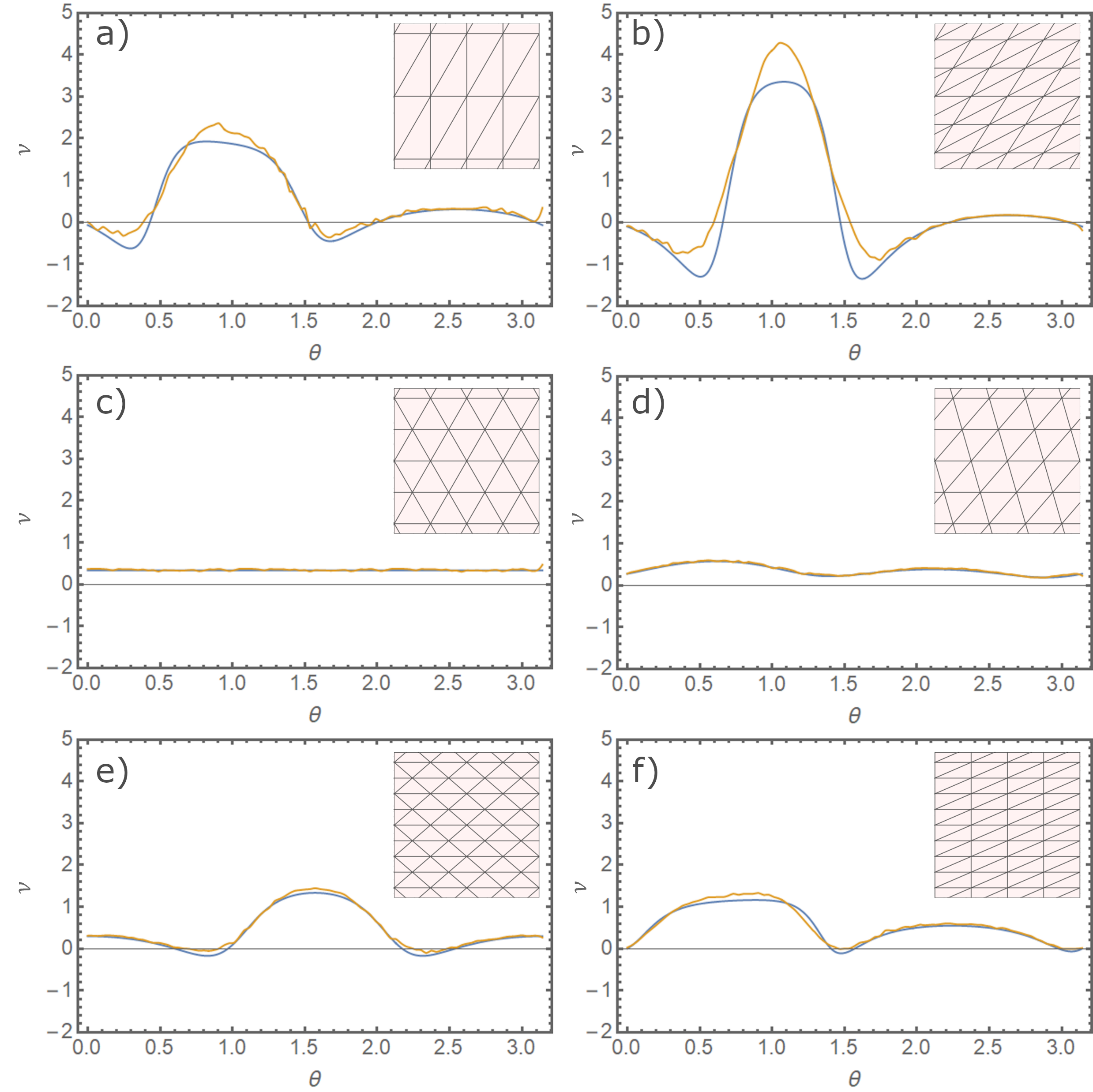}
	\caption{Ordered networks. Simulation (yellow) vs analytical estimation (blue) of Poisson's ratio as a function of the angle depending on different shape parameters $(\phi, \psi)$. a) $(\phi=2,\psi=2)$, b) $(3,0.9)$, c) $(1, 1)$, d) $(1.5,1)$, e) $(1,0.5)$, f) $(2,0.5)$. Insets- the spring network shape {at the initial orientation $\theta=0$. Other orientations are created by rotating the pattern, while keeping the ribbon orientation fixed.}  } \label{fig: numerics_vs_analytics}
\end{figure}

\subsubsection{Foam-like}
\begin{figure}
	\centering
	\includegraphics[width=\textwidth]{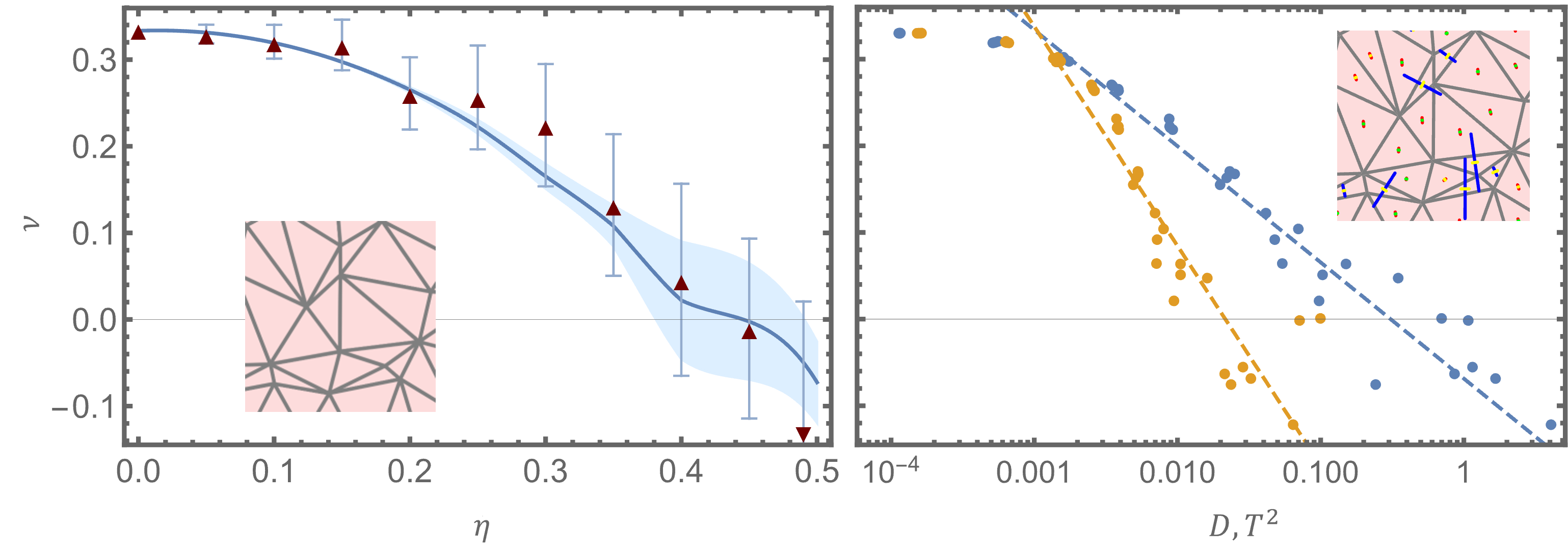}
	\caption{
	Random foam-like networks. Left - Comparison between theoretical calculation (solid line) and numerical simulation (triangles). Solid line represents the average of 7 different realizations of about 150 vertexes each (shaded region is typical deviation). Numerical simulations were done over 10 realizations of 676 vertexes each, error bars mark the deviations. Note that the last triangle is pointing down, indicating the average is beyond plot boundaries. Inset- an example of a $\eta =0.3$ realization. Right - Relation between Poisson's ratio, to the  response averages $D$ (yellow) and $T^2$ (blue) (see text for description) quantifying the network response. Points are the calculated results for different realizations, dashed lines - trend indicators. Inset - same inset as in Left, this time with principal components of $\frac{\delta g_{\mu\nu}}{\epsilon}$. Blue and red - elongation or contraction of major principal direction (respectively). Yellow and green - out-of-phase or in-phase response of minor principal direction relative to the major direction. Size of lines- relative size of principal response coefficients.}\label{fig: WOW!}
\end{figure}

Following \cite{Liu2019} we simulate a random, foam-like, network, exhibiting  an auxetic behavior at certain parameter range. Network is produced by {shifting} each vertex position of a regular triangular lattice, in a random direction by a fixed amount $0 < \eta < 0.5$. 
Calculation was done several times to average the results. Our results are consistent with the those in \cite{Liu2019} - $\nu$ decreases as a function of $\eta$ reaching $\nu=0$ at $\eta \simeq 0.46$ and reaching $\nu = -0.1$ when $\eta \rightarrow 0.5$. In the left side  of Fig. \ref{fig: WOW!} we  compare the simulation (discrete triangles) and the semi-analytical computation described in this paper {(smooth line).  All the springs are assumed to have the same spring constant in this case (which was conveniently chosen as $1$). As described in the methods section, simulation was done over a "strip" with about 680 vertexes and an aspect ratio of 4:1, mimicking a real experiment (a large aspect ratio is required in order to eliminate boundary effects). On the other hand the calculation introduced in this paper was done using a geometrical mean field (see methods and also appendix B \cite{supp}) which naturally eliminated the boundary effects expected in an exact solution. As a direct result of this choice, there was no need of a large network, and a square network with about 150 vertexes proved enough. The fact that such very different methods of calculation produce similar results points at the strength of this method.} 
 
{Additionally, the network response can be computed and visualized. {Ideally we would like to quantify the elements of $W^{\mu\nu\alpha\beta}_{(s)}$ as they are the basic quantities. However, for simplicity, we opted for quantifying the "local metric "response} $\delta g_{\mu\nu} = W^{\alpha\beta}_{\mu\nu}\Delta g_{\alpha\beta}$.  An example of the visualization may be found in inset of right image in Fig. \ref{fig: WOW!} {for the case of $\eta=0.3$}. Most notably, it is clear that the change in Poisson's ratio is related to the response of high aspect-ratio triangles that occur randomly via the process described. In these triangles one of the edges is exceptionally short. {Since all the spring have the same spring constant,} this translates to lower local rigidity value which intuitively means stronger response. To quantify the total response of a network, we calculated the eigenvalues of the response coefficients $\frac{\delta g_{\mu\nu}}{\epsilon}$ of each triangle, then averaged the determinant ($D = \det \frac{\delta g_{\mu\nu}}{\epsilon}$) and the square of the trace ($T^2  = \left(\Tr \frac{\delta g_{\mu\nu}}{\epsilon}\right)^2$)  (as the average of the trace by definition vanishes over the whole network). These are plotted on the right side of Fig. \ref{fig: WOW!}.}

\subsubsection{Hexagonal (honeycomb) network}
\begin{figure}
	\centering
	\includegraphics*[width=.67\textwidth]{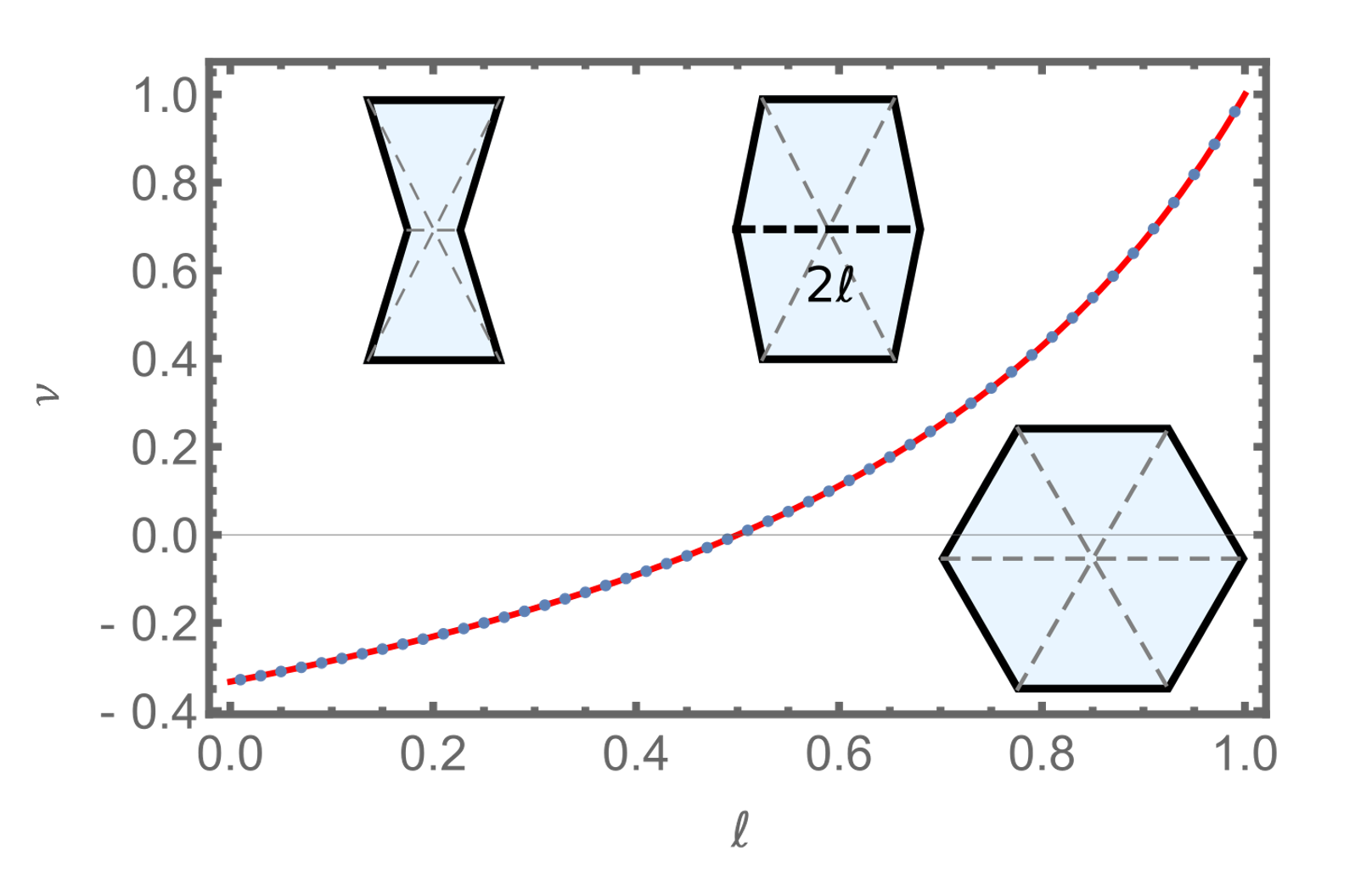}
	\caption{Hexagonal networks. Theoretical (solid line) and computational (points) results for a honeycomb made out of uniform hexagons with diameter $2\bar\ell$. To avoid singular expressions, each hexagon was divided to triangles (as indicated by dashed lines). Such that the added edges had negligible, but finite, rigidity ($k_{dashed}=10^{-3}k_{solid}$), insets (from left t right - re-entrant hexagon, general hexagon, regular hexagon)}\label{fig: Hex}
\end{figure}

We consider a honeycomb network, in which the basic hexagonal unit can vary between regular and a re-entrant hexagon, continuously with a diameter (distance between two opposing vertexes) of $2\bar\ell$ $0<\bar\ell<1$. (see inset in Fig. \ref{fig: Hex}). In such a case, Poisson's ratio is analytically given \cite{Gibson1982}
\begin{align}
	\nu(\bar\ell)&=-\frac{\left(2-4\bar\ell\right)\left(\bar\ell+1/2\right)}{3+4\bar\ell-4\bar\ell^2}
\end{align}

In order to calculate the elastic response, we use a triangularized hexagon, with a vertex at the center, and set the spring constant of the radial springs connecting the center with each corner of the hexagon to a very small value ($1/1000$'th of the peripheral springs). Without this, the original formulation become singular when the spring constant vanishes completely.  Results are shown in Fig. \ref{fig: Hex} with a very high degree of agreement between the analytical result and our formulation, despite the use of a large difference between springs constants, strengthening our approach.

\blu{\subsection{Young's modulus}
	In the previous sections we used a simplified requirement over the non-affine deformations $\delta g_{(s)\mu\nu}$'s. In fact, it a single constraint, for  the $N$ simplexes. However, a simple consideration reveals this constraint is not enough, as it leaves too many degrees of freedom compared to the original problem. \newline	
	Originally, in $d$ dimensions we have $n_V$ vertices each with $d$ degrees of freedom (total of $d\times n_V$ degrees of freedom). In this formulation, we have $n_s$ simplexes, each with a local metric of its own, which contains $d(d+1)/2$ independent components. Typically, $\frac{d(d+1)}{2} n_s > d n_v$. We thus require more constraints, these lie in the requirement that the length of an edge must be agreed upon by all adjacent local metrics (in two dimensions there are only two simplexes adjacent to each edge, but in higher dimension, there are more, depending on the coordination number). Additionally, we require that whatever the total volume of the deformed system is, it must be the same regardless of the use of the local metric $g_{(s) \mu\nu}$ or the average $g_{\mu\nu}$. In other words 	
	$$ V_{tot} = \sum_s{\sqrt{\frac{\det{g}}{\det{\bar{g}}}} V_s} = \sum_s{  \sqrt{\frac{\det{\left[g + g_{(s)} \right] }}{\det{\bar{g}}}} V_s} \simeq \sum_s{V_s \sqrt{\frac{\det{g}}{\det{\bar{g}}}}(1+ \Tr{\left[g^{-1}\delta g_{(s)}\right]})} = \simeq \sum_s{V_s \sqrt{\frac{\det{g}}{\det{\bar{g}}}}(1+ \Tr{ \left[\bar{g}^{-1}\delta g_{(s)}\right]})}$$
	Where $V_s$ is the original volume of the simplex. This means that we have an additional requirement that $$\Tr{ \left[\bar{g}^{-1}\delta g_{(s)}\right]} =0 $$ to leading order. \newline	
	Using these constraints, while formally more accurate, is rather cumbersome and difficult to implement. However,  as shown in this and previous sections, these can be successfully replaced with $\sum_s \delta g_{(s)\mu\nu} =0$ for the geometrical response, and that for the energy we require the weighted sum $\sum_s \frac{V_s}{\sqrt{\bar{g}}} \delta g_{(s)\mu\nu} =0$. The first requirement works for the geometrical response since less rigid regions contribute significantly to the total response (easier to deform) and are important (at least as evident by analysis in previous sections). However, since we have too many degrees of freedom, we are bound to reach an energy minimum which is unattainable physically. \newline	
	The second constraint, however, gives a better approximation of the energetic contribution (smaller, softer regions contribute less to the total), at the price of giving those regions insufficient weight and therefor miscalculating the correct geometric response. \newline	
	For practical use, we follow the exact same derivation as in the previous section, replacing the constraint and sums are turned to volume-weighted  sums, the expression look formally the same.
	\newline
	In any case, these consideration are relevant only for the disordered case. For an ordered lattice the calculation yields an exact result. The results for an ordered case are shown in Fig. \ref{fig: Youngs lattice}, and for foam-like in Figs. \ref{fig: Youngs} and \ref{fig: Bulk and shear disorder}.  Additionally, oftentimes in numerical calculation, rigidity is of lesser importance, as it can always be chosen or normalized to whatever value needed, however, the exact geometric response (i.e ratios between different components) cannot be normalized and is harder to reproduce. \newline	
	\begin{figure}
		\includegraphics[width=0.7\textwidth]{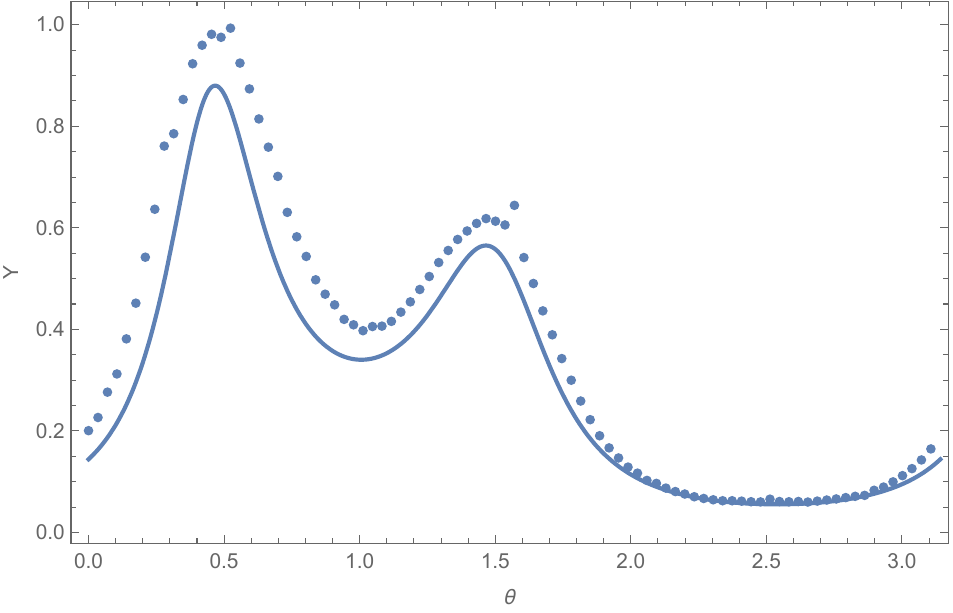}
		\caption{Angle dependent Young's Modulus as function of angle for triangular lattice with parameters $\phi=3$, $\psi=2$. Simulation results in dots, calculation in solid line. \label{fig: Youngs lattice}}	
	\end{figure}	
	\begin{figure}
		\includegraphics[width=0.7\textwidth]{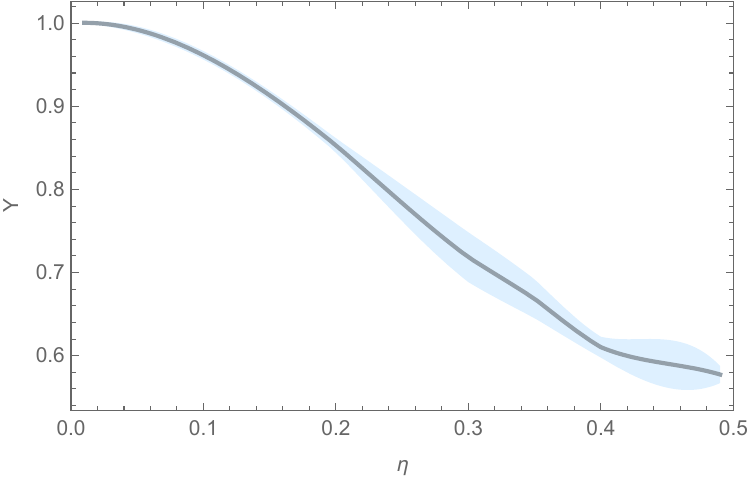}
		\caption{Young's modulus, correctly weighted. Solid lines are the average values over 10 different realizations for each $\eta$, shaded regions are indicate the spread of those values.\label{fig: Youngs}}	
	\end{figure}	
	\begin{figure}
		\includegraphics[width=0.7\textwidth]{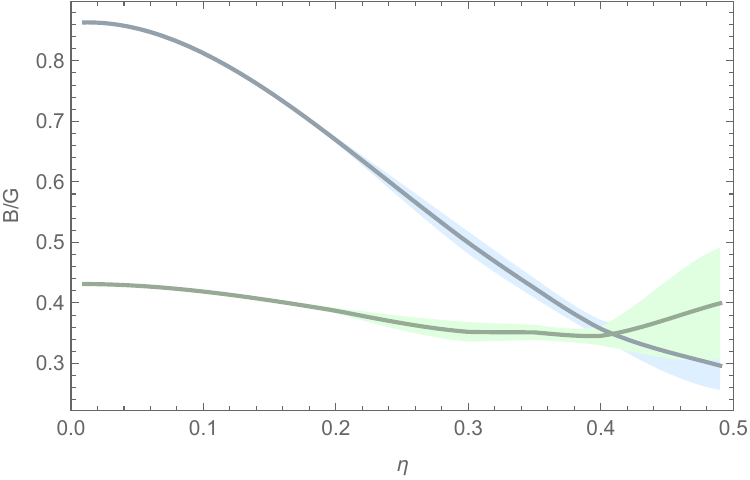}
		\caption{Bulk modulus (B, blue) and shear modulus (G, Green), derived from the corrected Young's modulus, and Poisson's ration as derived in the text. Correctly recovering results from \cite{Liu2019}. Solid lines are the average values over 10 different realizations for each $\eta$, shaded regions are indicate the spread of those values.\label{fig: Bulk and shear disorder}}	
	\end{figure}
}

{\section{Discussion}
	Here we  discuss the relevancy of assumptions made and the implications of their breakage on the coarse gaining process and the calculations made here. Additionally, we will exemplify why, under these assumption, there is no need to consider other non-flat geometries.
	\subsection{Assumptions}
	\begin{enumerate}
		\item 	The first assumption we made is that the reference lengths satisfy the triangle inequality. Mathematically speaking, if this is not satisfied, then the geometry cannot be described by a reference metric, as any metric space must satisfy this inequality. Physically, such a system will be highly frustrated, with residual stresses that are very large relative to the system and  typically cannot occur in nature (typical geometry is of system-size scale, rather than  element size - e.g. consider the shape of a leaf, the cells are much smaller than it's typical curvature). Possibly certain glassy systems may exhibit a few of such regions\blu{, but} this remains to be tested.  Theoretically, unless \blu{this assumption breaks in some small manner}, the derivation presented here cannot continue beyond Eq. \ref{eq: no ref metric}, and a different treatment is required.
		\item The second assumption made is that actual lengths are close to reference lengths.  \blu{Given that} our first assumption is valid\blu{, this} assumption when deformations are very localized and very large. This is  not a typical realistic case (materials tend to break on such scales). Considering a theoretical hyperelastic material, in which this assumption cannot be made, is an interesting problem in and of itself, especially from the mathematical point of view, as it involves averaging non-trivial functions of tensors.
		\item Finally, the last assumption is, essentially, simply the continuum limit. That is - there are many elements within a region enough to \emph{self average} so that $\bar{g}$ is well defined on the region. If variation of $\bar{g}$ occur on finite regions, then (as often the case is) finite size effects cannot be neglected. It is known \cite{Tsamados2009} that disordered systems converge only slowly to this limit, and as such finite size effects may become important even at relatively large scales (with respect to element size). The treatment of finite size effects is kept to a later opportunity, and is beyond the scope of this work.
	\end{enumerate}
	Note that for every point on a continuous medium, we may choose what are known as "normal coordinates" around that point. In these coordinates, the reference metric can be written as:
		\begin{align}\label{eq: metric normal coordinates}
		\bar{g}_{\mu\nu} =\delta_{\mu\nu} - \frac{1}{3} \bar{R}_{\mu\nu\rho\sigma}x^\rho x^\sigma + \cdots 
		\end{align}
		where $\bar{R}$ is the Riemann curvature tensor of the reference metric.
	Eq. \ref{eq: metric normal coordinates} means \blu{that} the deviation of the metric from euclidean one is quadratic around this point and heavily depends on the (intrinsic) curvature. In a continuous material, we can always choose a small enough region so that this correction is arbitrarily small. Hence, as long as we work with expressions which are covariant, it is enough to prove them locally, and general covariance will "take care of the rest". 
	\newline
	In our case, though, we do not deal with a continua. However, this is exactly the meaning of our last assumption: that elements are small enough and multitude, so that we may still work in a small enough region such that the correction term in Eq. \ref{eq: metric normal coordinates} is negligible. }
\blu{
\subsection{Context and relation to past works}
In the past, works have been focused on compatible and flat geometries, or close to such (aside of finite discrete points).  In that sense, they represent a family of measure zero among all possible elastic models, especially in relation to residually stressed systems.   In this aspect, the formulation presented here encompasses a significantly more general result which can be further generalized to include active stresses (such as in \cite{sheinman2012actively, scheibner2020odd}) which will likely couple non-trivially to curvature and residual stresses.\newline One way to study the coupling of active stresses in non-trivial geometry, is to derive the 3D description of active systems, and the dimensionally reduce the 3D model to an effective 2D or even quasi-1D as was done in \cite{Efrati2009,grossman2016elasticity,Levin2021}\newline
The coarse-graining approach introduced here diverges from standard coarse graining approaches \cite{scheibner2020odd, Seung1988,Zaccone2011} or Effective Medium (EM) approaches \cite{sheinman2012nonlinear} in  several important ways, despite some similarities.  First, the metric approach is inherently non-linear. Strain in this formulation is not a deformation gradient - it is a metric difference. This allows for treatment of relatively large geometric differences, even when a reference configuration cannot be defined (and therefore a deformation relative to it cannot be defined). In this aspect the derivation here diverges significantly from Bohr approach and more standard ways to coarse grain elasticity. Additionally, it is this non-linearity which at all allows dealing with non-trivial geometries and residually stressed ones at that as explained in the discussion of our assumptions.  \newline Second, in contrast to effective medium approaches, here we do not make analogies to ordered media (which is inherently flat and compatible), nor do we limit the calculation to a single mode of deformation. Similarities arise due to similar looking expansions of non-affine terms relative to the affine terms, but unlike EM methods, here we calculate explicitly the response of every element in the network, to every possible deformation. It would be very interesting to see how EM approaches might be used to derive completely analytic expressions, but these require some prior knowledge on the distribution of possible non-affine response, or otherwise, like in EM other self consistent condition.
}
\section{Conclusions}
In this work we give, to our knowledge, the first analytical derivation of the effective, coarse grained, elastic description of a general spring network. {This work generalizes a large body of work in different fields regarding the mechanical response of spring networks \cite{Martin1972,Miserez2022,Lemaitre2006,Zaccone2023,Schlegel2016,Cui2019,Zaccone2011,Seung1988,lloyd2007identification,plummer2020buckling,leembruggen2023computational,ostoja2002lattice,gelder1998approximate}. It expands this work in its generality, as it encompasses a much wider range of possible reference geometries which are not necessarily flat and indeed may not be even attainable within the confines of three dimensional euclidean space. Such systems are residually stressed (as there is no stress-free rest configuration) and are common in man made and especially biological systems that can grow and move.}

{Among the important aspects of this paper, is the formal derivation of the continuum limit of the formulation of incompatible elasticity, which was derived originally from a phenomenological aspect \cite{Efrati2009}.} 

Comparison  of computational results stemming from this derivation to known/numerical results shows a high degree of agreement.  Additionally, we identified and quantified the "non-affine" deformations, and have shown how they affect the resulting continuum elastic model, and related them to network structure. In systems such as granular media, these quantities play a role in local stress release by means of plastic deformations \cite{Tsamados2009}.
 
While derived over a spring-network, the results' derivation shown is relevant to many fields and systems in two ways. First the interaction between elements is almost always approximated as that of a simple spring, especially at small deformation. This is true for mechanical systems such as meta materials \cite{Rocks2017,Liu2019,Luo2023} and coarse grained mechanical models\cite{Bolander1998}, chemical systems such as self assemblies \cite{Kantor1987,Seung1988,Bailey2021,Underhill2004}, and many biological systems as well \cite{Moshe2018,Chen2014}. As such, the resulting theory, as is, is relevant to engineers, physicists, chemists, and biologist, and opens the possibility of rational design of materials.

Second, the coarse graining process described here can be generalized to other, more complicated interactions, and is not limited to point masses connected by linear springs. Nonlinearities can be addressed by using higher order terms (shape - related nonlinearities are already addressed by the usage of a metric description). {Adding angle - preferring interaction is also a relatively simple generalization that may be relevant even in polygonal networks (see appendix D for example)}. Formulated correctly it could also apply to complex molecules, cell-cell interactions, and to polymer-networks.  Activity may be involved in it as well. It would be very interesting to see how the activity \cite{scheibner2020odd,sheinman2012actively}, or non trivial network topology \cite{sheinman2012nonlinear}) may couple to geometry, possibly giving rise to additional geometrical terms, resulting in a richer and more complex theory.

Finally, the introduction of the new, $W$, quantities invites further investigation as to the nature of the solutions of Eq.\eqref{eq: non_affine_cental}, both analytically (possible through Effective Medium \cite{sheinman2012nonlinear} approaches), and numerically. It is known, for example, that the non-affine deformations {become substantially less significant at large scales (though strictly scale independent) }\cite{Tsamados2009}, this formulation may allow further insight into their scale dependence, and possible interaction with curvature.  Other usages would involve intelligent design - relating the required mechanical behavior to the non-affine deformations, and from that to the network structure.

\section{Methods}

We used our assumptions that $\bf \bar{g}$ is well defined on a large enough region to limit our numerical and analytical solution for the case $\bar{g}_{\mu\nu} = \delta_{\mu\nu}$, as we can always work in a locally flat frame.  This condition is sufficient as we want to isolate the effects of the non-trivial structure of the network itself, not the whole  (non uniform) mechanical response of a complex, possibly residually  stressed structure.  We compared the results of 3 test cases, ordered (non isotropic), foam like (following procedure in \cite{Liu2019}), and a honeycomb, despite the latter being strictly - non triangulated. The latter can be calculated analytically, rather than simulated. 

In parallel to simulation, for every network architecture we calculated $\tilde{A}^{\mu\nu\alpha\beta} $ and using it, we calculated the response to a hypothetical small strain $\epsilon$ by setting $g_{yy}= 1+ \epsilon$. Using the elastic equation \eqref{eq: Field equations}, and working in a geometric mean-field approximation, we solved the other terms $g_{xx}$ and $g_{xy}$ and calculate Poisson's ratio $\nu = - \frac{g_{xx}-1}{\epsilon}$ (see appendix B for detail \cite{supp}).

\subsection{Simulation}
The simulation was created for the purpose of this research. In each run we simulated a strip with length to width ratio of 4:1. A total of about $13 \times 13 \times 4 = 676$ vertexes, corresponding to about  $1000$ edges, depending on the exact details of each simulation. {Energy minimization was done using a simple gradient descent, implemented as the SciPy.Optimize.Minimize() class in Python}.

When creating a lattice, vertexes were positioned using the base vectors - 
\begin{align}
	v_1 =& \left(1,0\right) \\ \nonumber
	v_2 =& \left(\phi \frac{1}{2}, \psi \frac{\sqrt{3}}{2}\right)
\end{align}
where $0<\phi,\psi$ are the shear and elongation parameters, respectively, and are used to control the shape of the triangles.  $\psi =\phi = 1$ corresponds to an equilateral triangle, and any $\phi =1 $ is an isosceles triangle. The strip was created by keeping all vertexes whose coordinates satisfy $0\leq x \leq 13$ and $0\leq y \leq 52$ ("trimming").  Different orientations lattices were created by rotating the base vectors before trimming, so that the strip orientation remains constant, but the orientation of triangles relative to it changes.  

The set of vertexes was then used to create the list of edges via triangulation, and extracting the list of neighbors. The energy of each edge was directly calculated from the positions of its vertexes, using a simple spring energy.  In the simulation, the positions of the top  and bottom vertexes is held constant and all other vertexes are allowed to move in order to minimize the total energy.  

The two vertexes initially closest to $x=0, y=26$ and $x=13,y=26$ were identified to measure the strain between them $\delta = \frac{ \Delta x_{final}}{\Delta x_{intial}}$, where $\Delta x_{final}$ is the final $x$-coordinate difference between the two vertexes, and $\Delta x_{initial}$ is the initial difference.  
After setting the top vertexes at $y=52(1+\epsilon)$ ($\epsilon=0.01$), and letting the system to relax elastically,  Poisson's ratio was calculated via $\nu = - \delta/\epsilon$. And averaged over several simulations, if required (in the more stochastic simulations).

Simulating the foam-like structure, is stochastic in essence. We used the same initialization process, with the following differences. After generating a triangular lattice strip with $\phi=\psi=1$, and triangulation, we changed the position of each vertex by an amount $0< \eta< 0.5$ at a random direction, and used the resulting distances as the reference lengths of each vertex. We then followed the regular procedure by stretching the strip, and letting the system relax (with the reference lengths calculated just a moment before).

\subsection{Calculation through Eq.(20)}
A square patch was generated, independently of the simulation. Generation of the network itself was done similar to the way described in the simulation. However, once that calculated, instead of stretching the network we calculate $A^{\mu\nu\alpha\beta}$ using the $\{W_{(s)\,\lambda\tau}^{\mu\nu}\}$ which are calculated using Eq.\eqref{eq: non_affine_cental}. Poisson's ratio is calculated in the mean field approximation as described in appendix B \cite{supp}.

\section{Acknowledgements}
This research was partially supported by the Chaire Procédés \& Matériaux Innovants (Ecole Polytechnique - Saint Gobain)
D.G would like to thank Amos Grossman for his help, patience and useful discussions, and to Alessio Zaccone for pointing out important references. 

%


\clearpage
\appendix
\section{Supplemental Material}

\subsection{Appendix A:  Multi - index notation}
Here we write explicitly the expressions of the multi-index notation of the matrices $A^{SS'}$ and $B^{SS'}$ that appear in the text. One way to directly build them is to first write the matrices $W$ in terms of their independent terms (labeled $W^n$) and similarly for the matrix $A$.  Then  calculate the multiplication $M^{\mu\nu\alpha\beta}=A^{\mu\nu\rho\sigma}W^{\alpha\beta}_{\rho\sigma}$.  At this point we vectorize $W$, and use the same scheme to vectorize  $M$ (which has the same number of independent components as $W$), and $\delta A$. We now define
$A^{LL'} = \frac{\pd M_L}{\pd W_{L'}}$. 

E.g. in 2D - marking
\begin{align}
	A^{IJ}_s =&  \left(\begin{array}{cccc}
		A_s^{1111} & A_s^{1112} & A_s^{1211} & A^{1212} \\
		A_s^{1121} & A_s^{1122} & A_s^{1221} & A^{1222} \\
		A_s^{2111} & A_s^{2112} & A_s^{2211} & A^{2212} \\
		A_s^{2121} & A_s^{2122} & A_s^{2221} & A^{2222} \\
	\end{array}\right) 
	=  \left(\begin{array}{cccc}
		A_s^{1} & A_s^{2} & A_s^{2} & A_s^{3} \\
		A_s^{2} & A_s^{3} & A_s^{3} & A_s^{4}\\
		A_s^{2} & A_s^{3} & A_s^{3} & A_s^{4} \\
		A_s^{3} & A_s^{4} & A_s^{4} & A_s^{5} \\
	\end{array}\right)  \\ 
	W^{IJ}_s =&  \left(\begin{array}{cccc}
	W_{(s)\,11}^{11} & W_{(s)\,12}^{11} &W_{(s)\,11}^{12} & W_{(s)\,12}^{12} \\
	W_{(s)\,21}^{11} & W_{(s)\,22}^{11} &	W_{(s)\,21}^{12} & 	W_{(s)\,22}^{12} \\
	W_{(s)\,11}^{21} & W_{(s)\,12}^{21} & W_{(s)\,11}^{22} & W_{(s)\,12}^{22} \\
	W_{(s)\,21}^{21} & W_{(s)\,22}^{21} & W_{(s)\,21}^{22} & W_{(s)\,22}^{22} \\
	\end{array}\right) 
	=  \left(\begin{array}{cccc}
	W_s^{1} & W_s^{2} & W_s^{4} & W_s^{5} \\
	W_s^{2} & W_s^{3} & W_s^{5} & W_s^{6}\\
	W_s^{4} & W_s^{5} & W_s^{7} & W_s^{8} \\
	W_s^{5} & W_s^{6} & W_s^{8} & W_s^{9} \\
\end{array}\right)
\end{align}
Where repeating markings indicate that these elements are equal. Additionally the numbering of the $W_s$ matrix elements suggest a possible vectorization scheme (which is used in this text).

It now follows by direct calculation that after vectorization
\begin{align}
	A^{LL'}_s =&  \left(\begin{array}{ccc ccc ccc}
		A_s^{1} & 0  & 0 & 2 A_s^{2} & 0 & 0 & A_s^{3} & 0 & 0 \\
		0 & A_s^{1} & 0  & 0 & 2 A_s^{2} & 0 & 0 & A_s^{3} & 0  \\
		0& 0 & A_s^{1} & 0  & 0 & 2 A_s^{2} & 0 & 0 & A_s^{3}   \\
		A_s^{2} & 0  & 0 & 2 A_s^{3} & 0 & 0 & A_s^{4} & 0 & 0 \\
		0 & A_s^{2} & 0  & 0 & 2 A_s^{3} & 0 & 0 & A_s^{4} & 0  \\
		0& 0 & A_s^{2} & 0  & 0 & 2 A_s^{3} & 0 & 0 & A_s^{4}   \\
		A_s^{3} & 0  & 0 & 2 A_s^{4} & 0 & 0 & A_s^{5} & 0 & 0 \\
		0 & A_s^{3} & 0  & 0 & 2 A_s^{4} & 0 & 0 & A_s^{5} & 0  \\
		0& 0 & A_s^{3} & 0  & 0 & 2 A_s^{4} & 0 & 0 & A_s^{5}   \\
	\end{array}\right)
\end{align}
$\delta A^{LL'}_s$ has the exact same structure, but with $\delta A_s^n$ rather than $ A_s^n$
We can now finally define $A^{SS'}$ and $B^{SS'}$ (independently of dimension)-
\begin{align}
	A^{SS'} =&  \left[diag_n (A^{LL'}_s)\right]^{S,S} = \left(\begin{array}{cccc}
	A^{LL'}_1 & 0 & 0 &  \\
	0 & A^{LL'}_2 & 0 & \cdots  \\
	0 & 0 & A^{LL'}_3 &  \\
	& \vdots &  &\ddots
\end{array}\right)\\
B^{SS'} =& \frac{1}{n} \left[\delta A^{LL'}_s \otimes ones_n \right]^{S,S} =\frac{1}{n} \left(\begin{array}{cccc}
	\delta A^{LL'}_1 &  \delta A^{LL'}_2 & \delta A^{LL'}_3 &  \\
	\delta A^{LL'}_1 &  \delta A^{LL'}_2 & \delta A^{LL'}_3  & \cdots  \\
	\delta A^{LL'}_1 &  \delta A^{LL'}_2 & \delta A^{LL'}_3  &  \\
	& \vdots &  &\ddots
\end{array}\right)
\end{align}

In three dimensions the same logic follows. The matrices look different, as they are much larger.  

\subsection{Appendix B:  Mean field approximation}
Within a geometric mean field, $\bf{g}$ is assumed constant in space. We begin by considering one edge lies at coordinate $y=0$, 
the other one at coordinate $y=y_m $, and similarly there are edges at $x=0$ and $x=x_m$. 
At time zero we deform the network in real space  $\vec{R} =\left\{X(x,y),Y(x,y)\right\}$ so that $\left| Y(y=y_m) -Y(y=0)\right| = L$. We take this constraint into account by introducing an effective energy  using a Lagrange multiplier $\lambda$ which is the stress on the boundary causing the stretching of the tissue.
\begin{align}\label{eq:sudden strain}
	E_{ext} & = \int  \lambda \left(\left| \vec{R}(y=y_m) -\vec{R}(y=0)\right|-L\right) \dt x = \int \dt x \lambda \left(\left| \int\limits_0^{y_m} \pd_y\vec{R} \dt y\right|-L\right) \dt x \\ \nonumber
	& =  \int  \lambda \left(\sqrt{\int \int  \vec{R}'(y) \cdot \vec{R}'(y')  \dt y' \dt y }-L\right) \dt x .
\end{align}
Within the mean field approximation, $\vec{R}'(y) = \pd_y \vec{R}(y) = const.$ So that $\vec{R}'(y) \cdot \vec{R}'(y')= g_{22}$ we can write
\begin{align}
	E_{ext} & =\int  \lambda \left(\sqrt{y_m^2 g_{22} }-L\right) \dt x = \int  \lambda \left(y_m \sqrt{ g_{22} }-y_m \sqrt{G}\right) \dt x=  \int   {\lambda} y_m \left(\sqrt{ g_{22} }-\sqrt{G}\right) \dt x \\\nonumber
	&= \int  {\lambda} \left(\sqrt{ g_{22} }-\sqrt{G}\right) \dt y \dt x  
\end{align}
where the length scale of the stretched network is $L=\sqrt{G} y_m$. 

Taking the variation of the total energy $E= E_{el}+ E_{ext}$, where is $E_{el} = \int \|{\bf g}- {\bf \bar{g}}\|^2 \dt V$  is the elastic energy (and $\bar{g}_{\mu\nu}=\delta_{\mu\nu}$) with respect to $g$ and $\lambda$ we derive the following equations:
\begin{align}\label{eq:sud strain EOM}
	\sqrt{g_{22}} - \sqrt{G} =0  \\ \nonumber
	\sigma^{\mu\nu} = \sigma_{el}^{\mu\nu} + \frac{\lambda}{2 \sqrt{g_{22}}} \left(\begin{array}{cc}
		0 & 0\\
		0 & 1 
	\end{array}\right)
\end{align}
where $\sigma_{el}^{\mu\nu}$ is the elastic stres
In principle, in order to find the metric $g_{\mu\nu}$ minimizing the elastic energy we must find the stress so that on the boundaries $y=0$ and $y=y_m$, the stress balances the force.  In the mean field approximation the conditions at the boundary impose $\sigma^{\mu\nu}= 0$. Using $G=1+\epsilon$ we get the required expression.

\subsection{Appendix C: Ordered networks}
In an ordered network, calculation is fairly simple.  As $\tilde{A}^{\mu\nu\alpha\beta}= {A}^{\mu\nu\alpha\beta}= {A}^{\mu\nu\alpha\beta}_s$ is uniform and the non-affine coefficients are 0. 
Where, with a little abuse of notation $$A^{\mu\nu\alpha\beta}(x) = \langle A^{\mu\nu\alpha\beta} \rangle = \frac{1}{N_s(\Omega)}\sum_{s\in \Omega(x)} A^{\mu\nu\alpha\beta}_{(s)}=  \frac{1}{N_s(\Omega)} \sum_{s\in \Omega(x)} \sum_{e \in s} \frac{k_e \Delta x_e^\mu \Delta x_e^\nu \Delta x_e^\alpha \Delta x_e^\beta}{16 \bar\ell_e^2},$$ and $\Omega(x)$ is some region around the point, x.

For an ordered systems, it is enough to consider just a small, finite sum of a few simplex edges (in two dimension these are just triangles).  We used $k_e=1$, and thus $A^{\mu\nu\alpha\beta} = \bar\ell_1^2 \left(\cos\theta_1\right)^{a} \left(\sin\theta_1\right)^{b} + \bar\ell_2^2 \left(\cos\theta_2\right)^{a} \left(\sin\theta_2\right)^{b} +\bar\ell_3^2 \left(\cos\theta_3\right)^{a} \left(\sin\theta_3\right)^{b}$, where $a=8-\mu-\nu-\alpha-\beta$, $b=4-a$, $\bar\ell_1, \bar\ell_2, \bar\ell_3$ are the reference lengths of the edges, $\theta_1, \theta_2, \theta_3$ are the angles of the edges relative to the "1" direction, and $\mu,\nu,\alpha\beta \in (1,2)$. We thus solve for $g_{11}-1=- \nu (g_{22}-1)$.

We modeled each triangle as composed of the following vertexes (up tp a global rotation angle $\Theta$ of the triangles relative to the stretching direction) $$\{(0,0), (\phi /2,\psi\sqrt{3})/2,(1,0)\}$$
$\phi$ corresponds to "shearing" of the lattice unit cell, while $\psi$ corresponds to "elongation".  This results with the edge list $$\{ (1,0), (\phi/2,\psi \sqrt{3}/2), (1-\phi/2, -\psi \sqrt{3}/2) \}$$ and the reference lengths $\{1, \sqrt{\phi^2+3\psi^2}/2, \sqrt{(1-\phi)^2+ 3 \psi^2}/2\}$

The theoretical calculation results with a complicated expression that can be evaluated 
\begin{align}
	\nu =& -\frac{P(\phi,\psi,\Theta)} { 4 Q(\phi,\psi,\Theta)}\\ \nonumber P(\phi,\psi,\Theta) = & \cos (4 \theta ) \left\{27 \psi ^6-(\phi -2)^2 \phi ^2 [\phi^2 -2\phi +4]+9 \psi ^4 (\phi^2 -2\phi) -3 \psi ^2 [(\phi^2 -2\phi)  (\phi^2 -2\phi -20)-16]\right\} \\ \nonumber &+\sin (4 \theta )\left\{2 \sqrt{3} \psi  (\phi -1) \left[9 \psi ^4+6 \psi ^2 (\phi^2 -2\phi -2)+(\phi^2 -2\phi) (\phi^2 -2\phi +8)\right]\right\}-48 \psi ^2\\ \nonumber &+\left(\phi^2 -2\phi -3 \psi ^2\right) \left[9 \psi ^4+6 \psi ^2 (\phi^2 -2\phi) +(\phi^2 -2\phi)  (\phi^2 -2\phi +4)\right] \\ \nonumber
	Q(\phi,\psi,\Theta) = & 36 \psi ^4 \sin ^4(\theta )+2 \cos ^4(\theta ) \left[ 6 \psi ^2 (\phi^2 -2\phi)^2 +(\phi^2 -2\phi)^2  (\phi^2 -2\phi +4)+9 \psi ^4 (\phi^2 -2\phi +2)\right] \\ \nonumber &+6 \psi ^2 \sin ^2(\theta ) \cos ^2(\theta ) \left(9 \psi ^4+\phi ^4\right)+3 \psi ^2 \sin ^2(2 \theta ) \left[3 \psi ^2 (\phi^2 -2\phi +2)-2 (\phi -2) (\phi^2 -4\phi +2)\right] \\ \nonumber & -4 \sqrt{3} \psi  (\phi -1) \sin (\theta ) \cos ^3(\theta ) \left(9 \psi ^4+6 \psi ^2 (\phi^2 -2\phi) +(\phi^2 -2\phi)(\phi^2 -2\phi +8)\right)-48 \sqrt{3} \psi ^3 (\phi -1) \sin ^3(\theta ) \cos (\theta ).
\end{align}

For each $\phi$ $\psi$ values the results are then plotted as a function of $\Theta$ and are compared to simulation.

\subsection{Appendix D: Contribution of preferred angles to the local Elastic tensor }
In this section we add a local term that may break the symmetry inherent to $A_(s)$ when calculated just from a triangulated spring. We now assume that between neighboring edges, there is also a preferred angle $\Theta_{ek}$ between the $e$th and $k$th edges. This can be addressed by adding term in the energy:
\begin{align}
	E_{ang}= \sum_{\langle e,k \rangle} \frac{1}{2} k'_{e,k} \left(\theta_{e,k}-\Theta_{e,k}\right)^2
\end{align}
where $k'$ is the rigidity associated with deviation from $\Theta$ and $\theta$ is the actual angle. 

Similarly, we may rewrite $E_{ang}$ using metrics, as follows:

\begin{align}
	E_{ang} = \sum_{\langle e,k\rangle}  \frac{1}{2} k'_{ek} \left(\theta_{ek} -\Theta_{ek}\right)^2 =  \sum_{\langle e,k\rangle} \frac{k'_{ek}}{2} \left(\pi - \arccos\left( \frac{\Delta x^\mu_e \Delta x^\mu_k g_{\mu\nu}}{l_e l_k}\right) - \Theta_{ek}\right)^2,
\end{align}
where we used the definition of an angle on a curved surface, given the metric $g_{\mu\nu}$, and the fact that we orient the edges.  As for the angle $\Theta_{ek}$ - we begin by assuming there is, in some sense, a "locally flat" frame in which these are the inner angles of a triangle. In other words $\sum_{\langle e,k\rangle \in i} \left(\pi - \Theta_{ek}\right) = 2\pi$. Otherwise, this indicates the existence of a monopole defect, as is the case, e.g. in a triangle with only $\pi/2$ angles. In such a case, the curvature scale is  of order of the cell size, in which case the edges cannot be considered as "straight", and $\ell_e = \int_e \sqrt{g_{\mu\nu}  \frac{d x^\mu_e}{d s}  \frac{d x^\nu_e}{d s}} ds $, over the curve representing the edge, with a curve parameter $s$.

Thus we may write $\Theta_{ek} = \pi - \arccos\left(\frac{\Delta x^\mu_e \Delta x^\mu_k G_{\mu\nu}}{\ell^G_e \ell^G_k}\right)$, where $G_{\mu \nu}$ is a reference metric (not necessarily $\bar{g}_{\mu\nu}$) and ${\ell_e^G}^2 \simeq G_{\mu\nu}\Delta x_e^\mu \Delta x_e^\nu $ is the edge's length according to $G$. Note that $\Theta_{ek}$ do not uniquely define a metric- as angles are a scale free measure. Thus if $G_{\mu\nu}$ describes angles, any $G'_{\mu\nu} = a G_{\mu\nu}$ for a scalar $a$, describes the same set of angles as well. 
Therefore,we limit ourselves to the case where $G_{\mu\nu} = \bar{g}_{\mu\nu}$.  Otherwise, we may keep the use of a general  metric, $G_{\mu\nu}$, but we will have to constrain it somehow for uniqueness. 

We may now write:
\begin{align}
	E_{ang} = & \sum_{\langle e,k\rangle}  \frac{1}{2} k'_{ek} \left(\theta_{ek} -\Theta_{ek}\right)^2 =  \sum_{\langle e,k\rangle} \frac{k'_{ek}}{2} \left(\arccos\left(\frac{\Delta x^\mu_e \Delta x^\mu_k g_{\mu\nu}}{l_e l_k}\right) - \arccos\left(\frac{\Delta x_e^\mu \Delta x_k^\nu \bar{g}_{\mu\nu}}{\ell_e\ell_k}\right)\right)^2 \\ \nonumber
	=& \sum_{\langle e,k\rangle} \frac{k'_{ek}}{2} \left\{\arccos\left[\frac{\Delta x^\mu_e \Delta x^\mu_k g_{\mu\nu}}{l_e l_k} \frac{\Delta x^\mu_e \Delta x^\mu_k \bar{g}_{\mu\nu}}{\ell_e \ell_k}+ \sqrt{\left(1-\frac{\Delta x^\mu_e \Delta x^\mu_k g_{\mu\nu}}{l_e l_k}\right)\left(1-\frac{\Delta x^\mu_e \Delta x^\mu_k \bar{g}_{\mu\nu}}{\ell_e \ell_k}\right)}\right]\right\}^2
\end{align}

Assuming small deviations of $g_{\mu\nu}$ or $\bar{g}_{\mu\nu}$. We get
\begin{align}
	E_{ang} = \sum_{\langle e,k\rangle} \frac{k'_{ek}}{2} \frac{\left(\frac{\Delta x^\mu_e \Delta x^\mu_k g_{\mu\nu}}{l_e l_k}-\frac{\Delta x^\mu_e \Delta x^\mu_k \bar{g}_{\mu\nu}}{\ell_e \ell_k}\right)^2}{1- \left(\frac{\Delta x^\mu_e \Delta x^\mu_k g_{\mu\nu}}{l_e l_k}\right)^2}
\end{align}
Finally

\begin{align}
	E_{ang} &= \sum_{\langle e,k\rangle} \frac{k'_{ek}}{2 \left[1- \left(\frac{\Delta x^\mu_e \Delta x^\mu_k g_{\mu\nu}}{l_e l_k}\right)^2\right]}\left[\frac{\Delta x^\mu_e \Delta x^\nu_k}{\ell_e\ell_k} \left(g_{\mu\nu}-\bar{g}_{\mu\nu}\right) -\frac{\Delta x^\mu_e \Delta x^\nu_k g_{\mu\nu}}{\ell_e\ell_k}\left(\frac{l_e-\ell_e}{\ell_e} + \frac{l_k-\ell_k}{\ell_k}\right)\right]^2 \\ \nonumber
	&= \sum_{\langle e,k\rangle} \frac{k'_{ek}}{2 \left[1- \left(\frac{\Delta x^\mu_e \Delta x^\mu_k g_{\mu\nu}}{l_e l_k}\right)^2\right]}\left[\left( \frac{\Delta x^\mu_e \Delta x^\nu_k}{\ell_e\ell_k} -\frac{\Delta x^\rho_e \Delta x^\sigma_k \ g_{\rho\sigma}}{\ell_e\ell_k}\frac{\Delta x^\mu_e \Delta x^\nu_e}{2\ell_e^2} - \frac{\Delta x^\rho_e \Delta x^\sigma_k \ g_{\rho\sigma}}{\ell_e\ell_k}\frac{\Delta x^\mu_k \Delta x^ \nu_k}{2\ell_k^2} \right)\left(g_{\mu\nu}-\bar{g}_{\mu\nu}\right)\right]^2 \\ \nonumber
	&= \sum_{\langle e,k\rangle} \frac{k'_{ek}}{2 \left[1- \left(\frac{\Delta x^\mu_e \Delta x^\mu_k g_{\mu\nu}}{l_e l_k}\right)^2\right]}\left[\left( \frac{\Delta x^\mu_e \Delta x^\nu_k}{\ell_e\ell_k} -\frac{\Delta x^\rho_e \Delta x^\sigma_k \bar{g}_{\rho\sigma}}{\ell_e\ell_k}\frac{\Delta x^\mu_e \Delta x^\nu_e}{2\ell_e^2} - \frac{\Delta x^\rho_e \Delta x^\sigma_k \bar{g}_{\rho\sigma}}{\ell_e\ell_k}\frac{\Delta x^\mu_k \Delta x^ \nu_k}{2\ell_k^2} \right)\left(g_{\mu\nu}-\bar{g}_{\mu\nu}\right)\right]^2 \\ \nonumber
	&= \sum_{\langle e,k\rangle} \frac{k'_{ek}}{2 \left(1- \cos^2(\pi-\Theta_{ek}) \right)}\left[\left( \frac{\Delta x^\mu_e \Delta x^\nu_k}{\ell_e\ell_k} -\frac{\cos(\pi-\Theta_{ek})}{2}\frac{\Delta x^\mu_e \Delta x^\nu_e}{\ell_e^2} - \frac{\cos(\pi-\Theta_{ek})}{2}\frac{\Delta x^\mu_k \Delta x^ \nu_k}{\ell_k^2} \right)\left(g_{\mu\nu}-\bar{g}_{\mu\nu}\right)\right]^2 
\end{align}
We may therefor mark 
\begin{align}
	A_{2,(s)}^{\mu\nu\alpha\beta} & =  \sum_{\langle e,k\rangle}  \frac{k'_{ek}}{2\left(1-\cos^2\Theta_{ek}\right)} \times  \\ \nonumber & \left( \frac{\Delta x^\mu_e \Delta x^\nu_k}{\ell_e\ell_k} +\frac{\cos\Theta_{ek}}{2}\frac{\Delta x^\mu_e \Delta x^\nu_e}{\ell_e^2} + \frac{\cos\Theta_{ek}}{2}\frac{\Delta x^\mu_k \Delta x^ \nu_k}{\ell_k^2} \right)\left( \frac{\Delta x^\alpha_e \Delta x^\beta_k}{\ell_e\ell_k} +\frac{\cos\Theta_{ek}}{2}\frac{\Delta x^\alpha_e \Delta x^\beta_e}{\ell_e^2} + \frac{\cos\Theta_{ek}}{2}\frac{\Delta x^\alpha_k \Delta x^ \beta_k}{\ell_k^2} \right)
\end{align}

This is term in not completely symmetric, and can be shown to drive the local Poisson's ratio to negative values. The reason is very simple - this term constrains only the angles, hence elongating in all directions does not change the energy. By adding it to the local elastic tensor calculated in the text, and following the same derivation we get the same expressions, the only difference would be in computing the local values, as there is now an additional contribution.

\subsection{Appendix E: Foam-like networks}
In this section we add figures of the geometry of the different networks of the foam like structures. All the left and right images are the same, the right images are all colored according to the triangles perimeter to area ratio (normalized to the same scale) to visualize better the the high aspect ratio triangles.

\begin{figure}[!h]
	\centering
	\includegraphics[width=\textwidth]{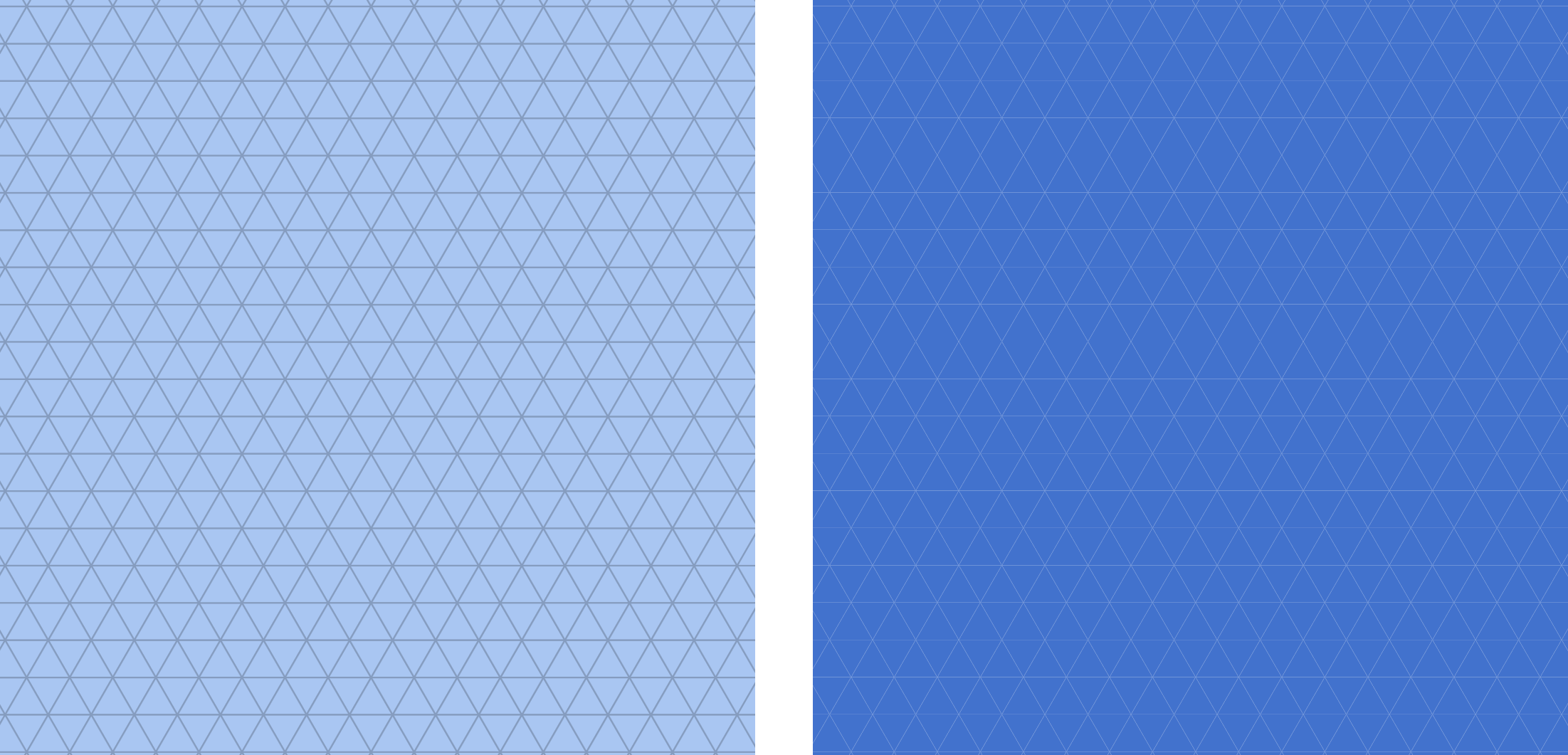}
	\caption{Foam-like network with $\eta =0 $ (regular lattice)}
\end{figure}
\begin{figure}[!h]
	\centering
	\includegraphics[width=\textwidth]{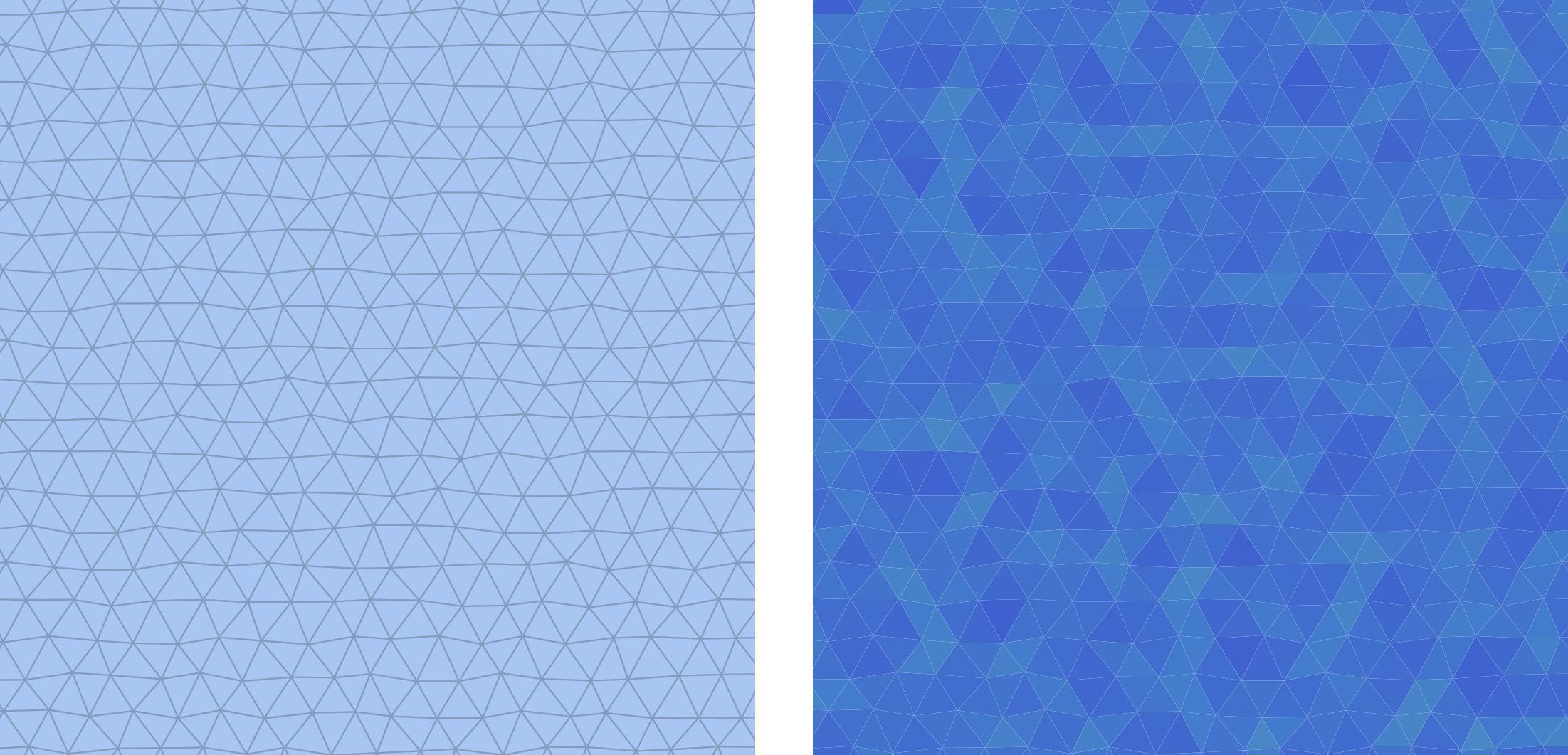}
	\caption{Foam-like network with $\eta =0.1 $ }
\end{figure}
\begin{figure}[!h]
	\centering
	\includegraphics[width=\textwidth]{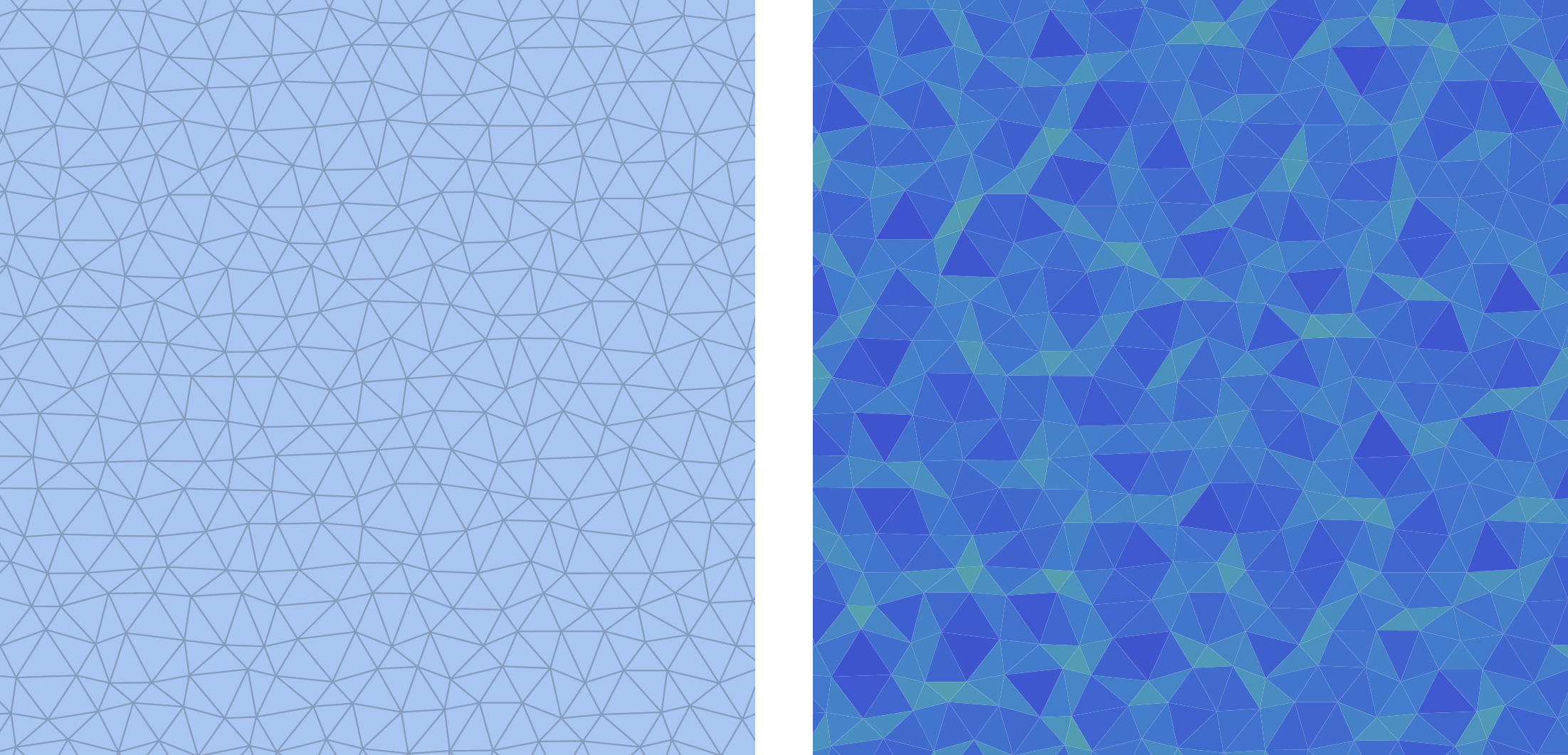}
	\caption{Foam-like network with $\eta =0.2 $ }
\end{figure}
\begin{figure}[!h]
	\centering
	\includegraphics[width=\textwidth]{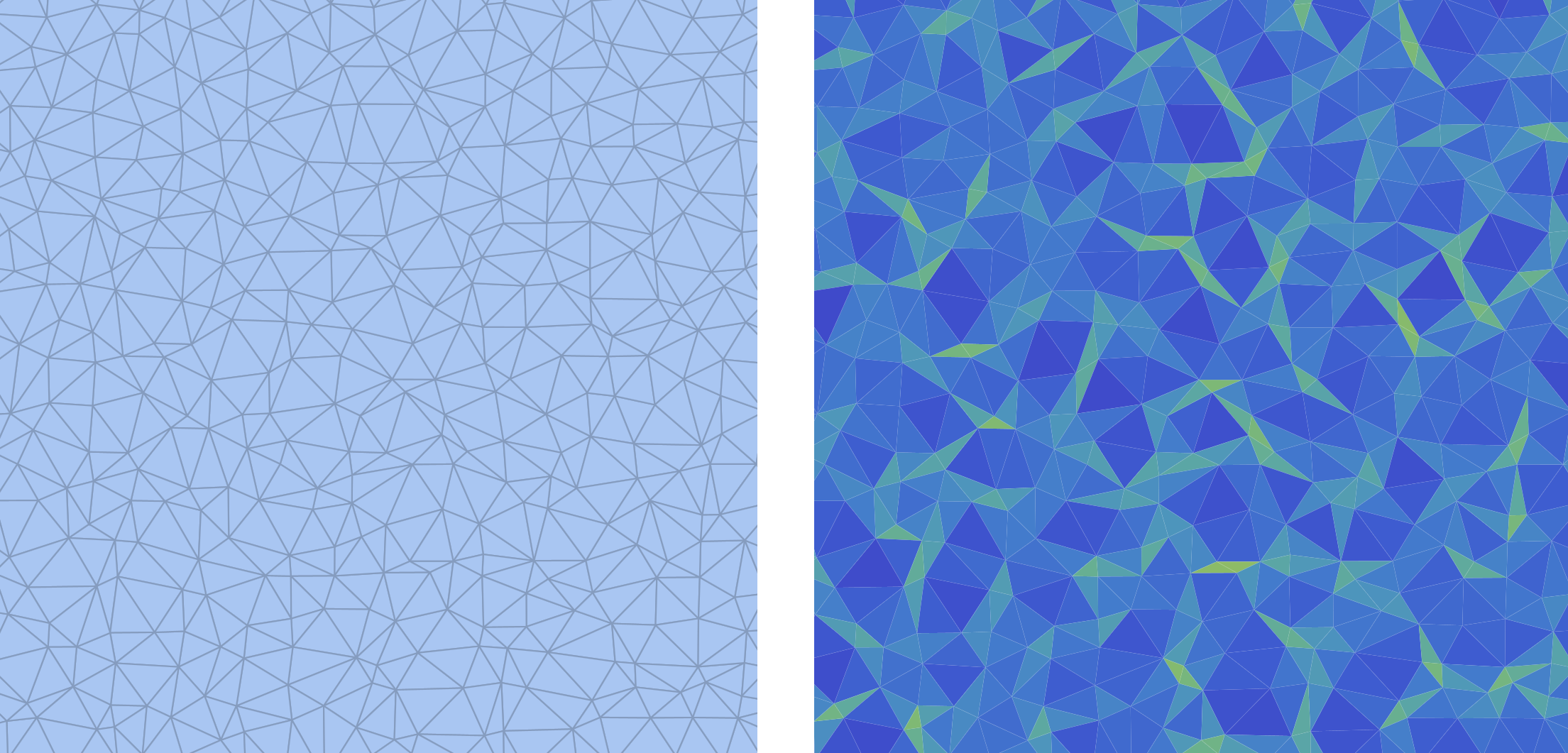}
	\caption{Foam-like network with $\eta =0.3 $ }
\end{figure}
\begin{figure}[!h]
	\centering
	\includegraphics[width=\textwidth]{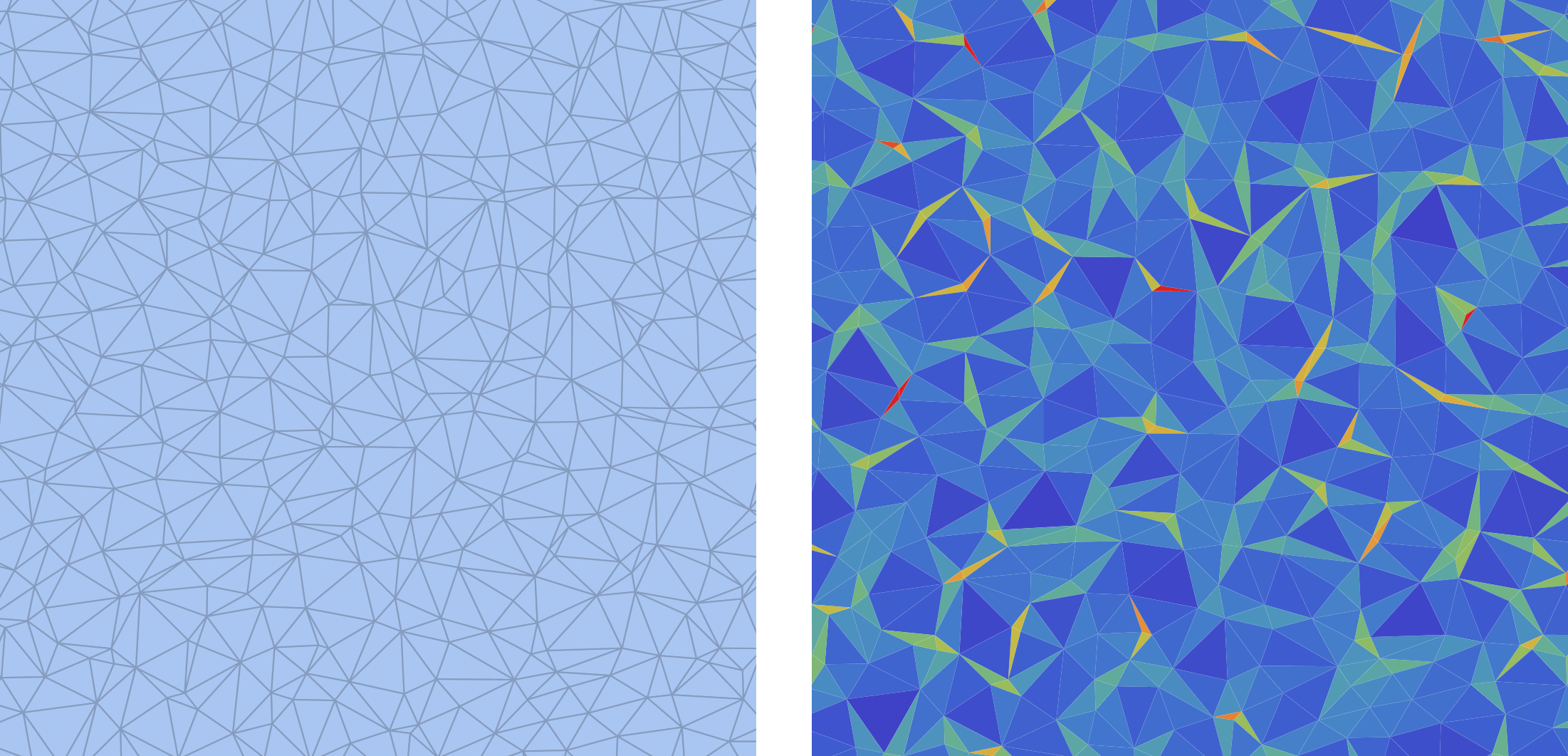}
	\caption{Foam-like network with $\eta =0.4 $ }
\end{figure}
\begin{figure}[!h]
	\centering
	\includegraphics[width=\textwidth]{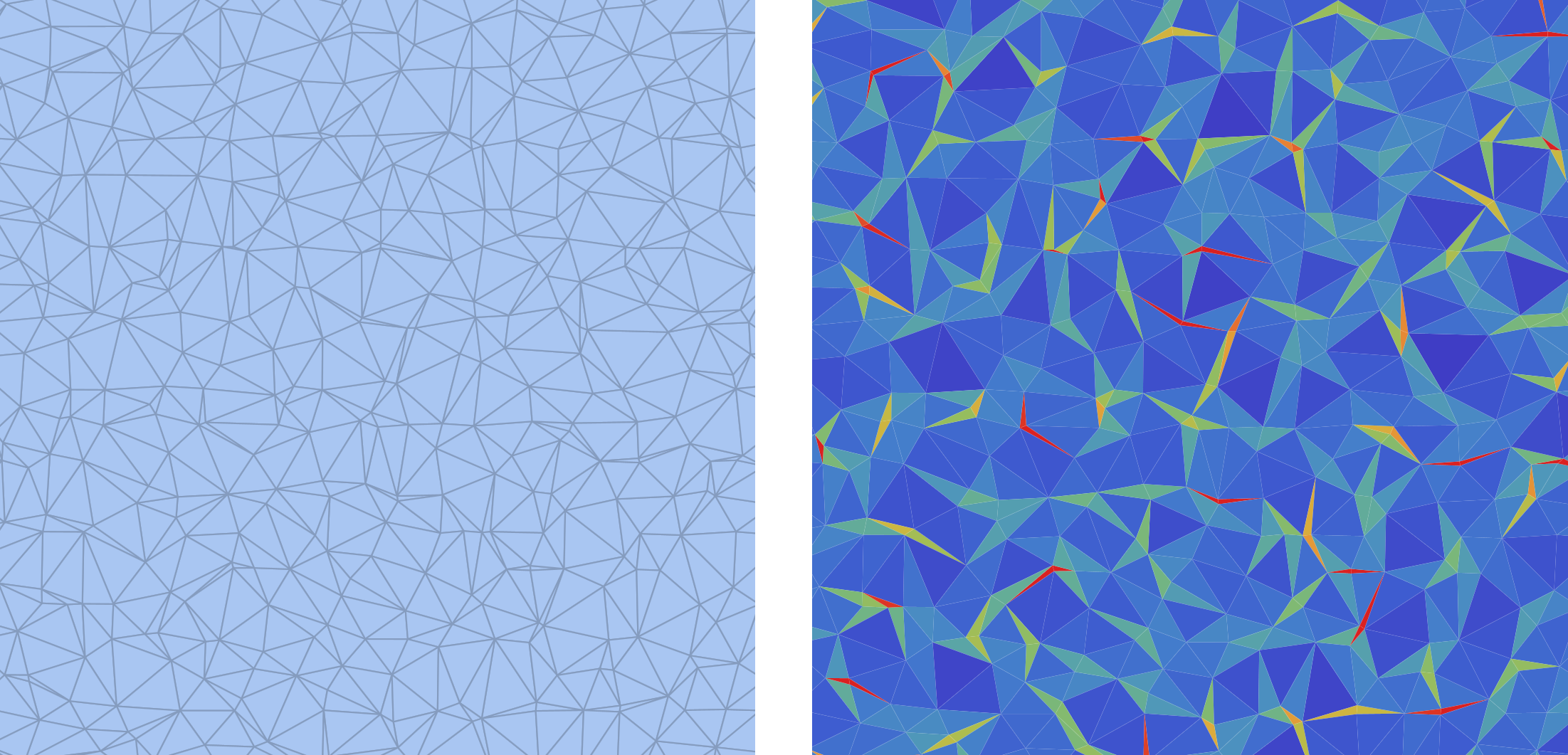}
	\caption{Foam-like network with $\eta =0.45 $ }
\end{figure}
\begin{figure}[!h]
	\centering
	\includegraphics[width=\textwidth]{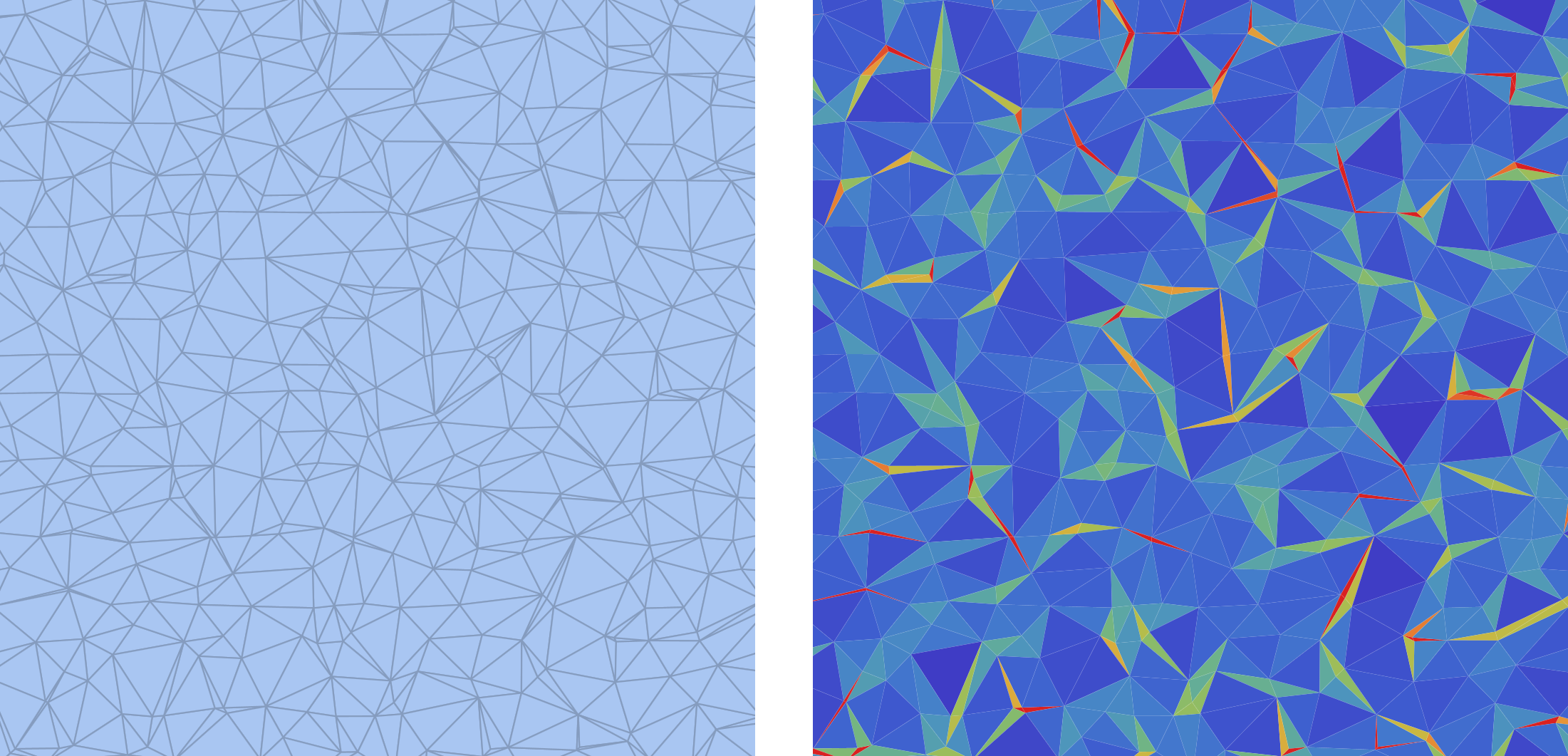}
	\caption{Foam-like network with $\eta =0.49 $ }
\end{figure}

 \clearpage

\subsection{Appendix E: Convergence}

\begin{figure}[!h]	\includegraphics[width=0.7\textwidth]{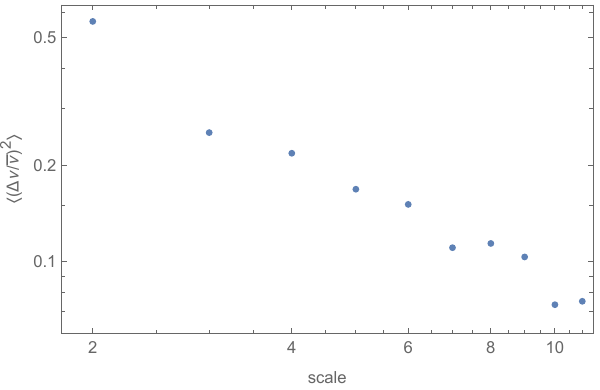}
	\caption{Standard deviation of the relative error of Poisson's ratio, for the case of $\eta=0.3$, $\delta \nu_{i,n}$. Showing a clear power convergence, similar to \cite{Tsamados2009}. 
		\label{fig: convergence}}
\end{figure}

We calculated the rate of convergence, i.e the importance of the non-affine displacements relative to the scale of the system. To quantify this, we generated 10 networks topologies, and kept only a portion of the original networks at different scales (scale from 2 (initial triangle size) to 11).  We then calculate the networks' Poisson's ratio $\nu_{i,n}$, for each topology $n$ and each scale $n$. We then calculated $\bar{\nu}_n$ , the average at each scale, and then we calculated standard deviation of $\delta \nu_{i,n} = \left(\nu_{i,n} - \bar{\nu}_n\right)/\bar{\nu}_n$. This quantifies the importance of a specific realization at a given scale.  The result are given in figure \ref{fig: convergence}.

 \end{document}